# Memory-corrected quantum repeaters with adaptive syndrome identification

Alena Romanova  [*] and Peter van Loock  [†]

*Institute of Physics, Johannes Gutenberg University Mainz, Staudingerweg 7, 55128 Mainz, Germany*

We address the challenge of incorporating encoded quantum memories into an exact secret key rate analysis for small and intermediate-scale quantum repeaters. To this end, we introduce the check matrix model and quantify the resilience of stabilizer codes of up to eleven qubits against Pauli noise, obtaining analytical expressions for effective logical error probabilities. Generally, we find that the five-qubit and Steane codes either outperform more complex, larger codes in the experimentally relevant parameter regimes or have a lower resource overhead. Subsequently, we apply our results to calculate lower bounds on the asymptotic secret key rate in memory-corrected quantum repeaters when using the five-qubit or Steane codes on the memory qubits. The five-qubit code drastically increases the effective memory coherence time, reducing a phase flip probability of 1% to 0.001% when employing an error syndrome identification adapted to the quantum noise channel. Furthermore, it mitigates the impact of faulty Bell state measurements and imperfect state preparation, lowering the minimally required depolarization parameter for nonzero secret key rates in an eight-segment repeater from 98.4% to 96.4%. As a result, the memory-corrected quantum repeater can often generate secret keys in experimental parameter regimes where the unencoded repeater fails to produce a secret key. In an eight-segment repeater, one can even achieve nonvanishing secret key rates up to distances of 2000 km for memory coherence times of $t_c = 10$ s or less using multiplexing. Assuming a zero-distance link-coupling efficiency $p_0 = 0.7$, a depolarization parameter $\mu = 0.99$, $t_c = 10$ s, and an 800 km total repeater length, we obtain a secret key rate of 4.85 Hz, beating both the unencoded repeater that provides 1.25 Hz and ideal twin-field quantum key distribution with 0.71 Hz at GHz clock rates.

## I. INTRODUCTION

Quantum repeaters are essential in enabling long-distance transmission of quantum information, which is otherwise hindered by photonic losses due to fiber attenuation. The transmission probability in optical fibers decreases exponentially with their length. For instance, over a distance of 100 km, on average, only one of one hundred photons survives the journey [1]. To counteract photonic losses, a quantum repeater [1,2] subdivides the communication distance into several segments with intermediate memory stations. One well-known application is quantum key distribution, which allows two parties to share a secret binary string and guarantees unconditional long-term security founded in the laws of physics, namely, the no-cloning theorem and quantum state collapse upon measurement [2]. On the contrary, classical post-quantum cryptographic systems rely on the complexity of certain mathematical tasks and are merely conjectured to be secure against quantum attacks [1]. In particular, long-term security cannot be guaranteed in this case.

Preserving entanglement is crucial to generate a secret key or to determine the presence of an eavesdropper [3]. However, under realistic circumstances, ambient noise or decoherence can severely corrupt the transmitted quantum state. Therefore quantum error detection and correction are indispensable tools to mitigate these effects, whereby stabilizer codes [4,5] will be our focus. Observing entanglement between single spins and photons has been achieved on various physical platforms [6], including trapped ions [7,8], neutral atoms [9–11], nitrogen-vacancy centers [12], or quantum dots [12,13]. Moreover, asynchronous photonic Bell state measurements, a crucial component of quantum repeaters, using a single-spin solid-state quantum memory have been successfully demonstrated [14]. Therefore single spins show great promise as a potential quantum memory. The results presented here directly apply to such stationary spin qubits but are not limited to them.

We extend the exact secret key rate analysis of Ref. [15] to memory-encoded quantum repeaters. Their developed theoretical model is purely analytical including state preparation errors, depolarizing errors for the Bell state measurements required for entanglement swapping, and dephasing errors accounting for the storage in quantum memories. For growing segment numbers, it was found that the secret key rate is particularly sensitive to faulty state preparation and Bell measurements, and the question was raised whether this hurdle could be overcome by employing encoded quantum memories, so by using a memory-corrected, also known as a second-generation, quantum repeater [16].

*Contact author: alena.romanova@uibk.ac.at
†Contact author: loock@uni-mainz.de

The idea to encode the memories of a quantum repeater was introduced in Ref. [17], where three and five-qubit repetition codes were investigated, generalizing to small Calderbank-Shor-Steane (CSS) codes. Placing a quantum memory station every 10 km, their repeater protocol with small CSS codes can increase communication distances to $10^3$–$10^6$ km and maintain a key generation rate above 100 bits per second using 30 to 150 qubits per station. Compared to the codes that we will utilize, this is rather resource-demanding. Later research has also investigated numerically repetition codes [18–20], for instance, applied to nitrogen-vacancy center memories [18,19] including memory decoherence with an approximation technique [19]. A surface code-based approach to quantum communication networks is presented in Ref. [21], describing how to extend the total transmission distance beyond 1000 km. However, for this, the author assumes a much higher single-link transmission probability, compared to what we will consider, and a lower qubit error probability, also requiring more than two hundred intermediate nodes. Other schemes rely on directly encoding the photon, the carrier of quantum information, in what is referred to as third-generation quantum repeaters [22–30].

Applying noise to stabilizer states results in mixed states that cannot, in general, be described as efficiently within the stabilizer formalism [4,5] anymore. In some special cases that, however, are not directly applicable in the context of our work, procedures have been established, for instance, for stabilizer mixed states [31] or for stabilizer states whose dimension decreases due to a series of Pauli measurements [32]. We shall treat ambient noise by extending and adapting the methods presented in Ref. [33] that translate Pauli noise channels, a superset of the depolarizing and the dephasing channels, into effective logical Pauli noise channels. Although their formalism applies to general stabilizer codes, the authors only consider repetition codes and the five-qubit code, potentially due to the computational complexity associated with density matrices of exponentially increasing size. Instead of working in the state picture as in Ref. [33], we shall utilize the stabilizer formalism, in particular, the binary check matrix of a stabilizer code [5]. As a consequence, we will be able to consider the effects of single-qubit noise on codes of up to eleven qubits and even generalize our model to two-qubit loss.

Importantly, we make use of an adaptive syndrome identification, where the mapping of the measured error syndromes onto the errors depends on the relevant quantum noise channel that is present in the corresponding part of the repeater protocol. For the five-qubit code, we exploit that it is possible to associate all fifteen error syndromes with phase flip errors on one and two qubits. This leads to an outstanding resilience against dephasing noise, reducing a phase flip probability of 1% by three orders of magnitude to 0.001%. Our results, in particular, the effective logical error parameters, may be insightful not only in the context of quantum communication but also for general quantum computing tasks, suggesting which stabilizer codes may provide the best protection of quantum information against common types of noise.

Provided the physical quantum noise channel remains unaltered on the logical level, albeit has modified error probabilities, one can directly insert the obtained error parameters into the secret key rate analysis of Ref. [15]. Our findings indi-cate that the five-qubit code and the Steane code are promising quantum memory encoding candidates since they increase the resilience against depolarizing and dephasing noise in the experimentally relevant parameter regime of low error probabilities and either outperform more complex codes or require fewer physical resources. Moreover, we demonstrate that encrypting quantum memories with these two stabilizer codes significantly extends the achievable distances for quantum key distribution and increases the rate of transmitted secret keys. Similar to Ref. [15], we will primarily focus on smaller quantum repeaters of up to eight segments, but we shall also briefly consider larger repeaters with hundreds or thousands of segments for a total distance of 800 km. This is interesting because, in this many-segment scenario, it is mainly the accumulation of errors from faulty initial states and gates that leads to a small or even zero secret key fraction, preventing the repeater from reaching secret key rates near the local clock rates as determined by the speed of the light-matter interfaces (typically of the order of MHz). Incorporating quantum error correction codes for the memory qubits may allow for suppressing the negative impact of these faulty elements.

For sufficiently good experimental parameters, our scheme outperforms not only the unencoded scheme but also ideal twin-field quantum key distribution [34] where an untrusted middle station is placed between the two end-node users and high source clock rates of GHz order can be used since there is no need for two-way classical communication. Nonetheless, an advantage of memory-assisted schemes is that we can scale a quantum repeater to more than two segments [35]. Additionally, the distribution of quantum states allows not only for secret key extraction as direct-transmission or twin-field quantum key distribution but also for distributed quantum computing applications.

The paper is structured as follows. In Sec. II, we briefly review the stabilizer formalism, focusing on stabilizer codes. Next, in Sec. III, we discuss the exact rate analysis from Ref. [15] and introduce our check matrix model that translates physical Pauli noise channels to the logical level after quantum error correction has been performed. The effective logical noise channels for various stabilizer codes of up to eleven qubits are then analyzed in Sec. IV, and the results applied to calculate secret key rates in memory-corrected quantum repeaters. Our findings are summarized and discussed in the final Sec. V.

## II. THEORETICAL BACKGROUND

### A. The stabilizer group

To define stabilizer codes [4,5], we consider the stabilizer group $S$, a subgroup of the $n$-qubit Pauli group

$$G_n = \{\pm I, \pm iI, \pm X, \pm iX, \pm Y, \pm iY, \pm Z, \pm iZ\}^{\otimes n},$$

where $I$ is the identity matrix, and $X, Y,$ and $Z$ are the Pauli matrices. The code space $V_S$ is defined as the vector space spanned by the quantum states, which are eigenvectors of all elements of $S$ with eigenvalue $+1$. In order to not stabilize the trivial vector space, we require all elements of $S$ to commute and that $-I \notin S$.

A compact description of the stabilizer group $S$ can be obtained via its generators, a set of elements $g_1, \ldots, g_l \in G_n$

such that any element of $S$ can be written as a product of the generators, which is denoted by $S = \langle g_1, \ldots, g_l \rangle$. Furthermore, we want our generators to be independent meaning that we cannot remove any generator $g_i$ without making the stabilizer group smaller, so for all $i \in \{1, \ldots, l\}$

$$\langle g_1, \ldots, g_{i-1}, g_{i+1}, \ldots, g_l \rangle \neq \langle g_1, \ldots, g_l \rangle.$$

Such $l = n - k$ independent commuting generators of a stabilizer group $S \subset G_n$, not containing $-I$, stabilize a vector space of dimension $2^k$, so a code space of $k$ logical qubits.

A particularly useful way to represent a stabilizer group $S = \langle g_1, \ldots, g_l \rangle$ is given by the $l \times 2n$ check matrix. It is constructed by identifying each generator $g_i$ with a matrix row $i$ via a group homomorphism $r$ from $G_n$ with multiplication to binary vectors of length $2n$ with addition modulo two:

$$e^{i\phi} X_1^{x_1} Z_1^{z_1} \otimes \cdots \otimes X_n^{x_n} Z_n^{z_n} \mapsto r(x_1, \ldots, x_n, z_1, \ldots, z_n),$$

where $\phi \in \{0, \frac{\pi}{2}, \pi, \frac{3}{2}\pi\}$ and $r(g_1 g_2) = r(g_1) + r(g_2)$. To represent a Pauli element $g \in G_n$ in terms of such a binary row vector, we use the notation $r(g) \in \mathbb{Z}_2^{2n}$ and define the check matrix as

$$\mathcal{C} = \begin{bmatrix} r(g_1) \\ \vdots \\ r(g_l) \end{bmatrix}.$$

Introducing the $2n \times 2n$ matrix

$$\Omega = \begin{bmatrix} 0_{n \times n} & I_{n \times n} \\ I_{n \times n} & 0_{n \times n} \end{bmatrix},$$

one can show that two elements of the Pauli group, $g$ and $g'$, commute if and only if $r(g)\Omega r(g')^T = 0$ and, correspondingly, they anticommute if $r(g)\Omega r(g')^T = 1$.

### B. Stabilizer codes

An $[n, k]$ stabilizer code, denoted by $C(S)$, consists of the codewords in $V_S$, a vector space stabilized by a group $S = \langle g_1, \ldots, g_{n-k} \rangle \subset G_n$ with $-I \notin S$ and $g_1, \ldots, g_{n-k}$ being independent and commuting generators. To each stabilizer code, we associate logical Pauli operators $X_1^L, \ldots, X_k^L \in G_n$ and $Z_1^L, \ldots, Z_k^L \in G_n$.

Error correction is achieved by first measuring all stabilizer generators $g_1, \ldots, g_{n-k}$. The results define the error syndrome, which is the outcome of a measurement determining what error, if any, has occurred. The syndrome reveals only whether an error has occurred but does not allow inferring anything about the state being protected. If a Pauli error $E$ anticommutes with a given stabilizer $s \in S$, so $\{s, E\} = 0$, we can detect the error by measuring $s$ since for $|\psi\rangle \in V_S$ it holds that

$$sE|\psi\rangle = -Es|\psi\rangle = -E|\psi\rangle.$$

So $E|\psi\rangle$ is an eigenvector of $s$, however, the eigenvalue has changed from $+1$ to $-1$. If instead the error $E$ commutes with $s$, $[s, E] = 0$, the eigenvalue $+1$ is unaltered. Hence, the error syndrome can be restated as the $n - k$ binary vector $f(E) = (f_{g_1}(E), \ldots, f_{g_{n-k}}(E))$, whereby for $s \in S$:

$$f_s(E) = \begin{cases} 0 & \text{if } [s, E] = 0 \\ 1 & \text{if } \{s, E\} = 0 \end{cases}.$$

Using the check matrix of a stabilizer code, the error syndrome becomes $f(E)^T = \mathcal{C}\Omega r(E)^T$. A set of errors $\{E_j\} \subset G_n$ acting on a stabilizer code $C(S)$ is correctable if

$$E_j^\dagger E_k \notin Z(S) \backslash S \quad \forall\ j, k,$$

where $Z(S) \supset S$ denotes the centralizer of $S$, so the set of elements that commutes with all members of $S$. This condition implies that an error that commutes with all stabilizer group generators but is not in the stabilizer group introduces a logical error and that two distinct correctable errors either have different error syndromes or their product leaves the code space invariant. Recovery is then accomplished by applying any Pauli error that has the measured syndrome.

The weight of an error $E \in G_n$ is the number of nonidentity terms in the tensor product. For example, the weight of $Y_1 X_3 Z_4$ is three. The distance $d$ of a stabilizer code $C(S)$ is the smallest weight of an element in $Z(S) \backslash S$. A code with distance $d$ of at least $2t + 1$ can correct arbitrary errors on up to $t$ qubits [5] and is usually denoted by $[n, k, d]$.

### III. METHODS

We employ the secret key rate analysis from Ref. [15] which we introduce in Sec. III A. In that reference, the secret key rate of repeater chains of various sizes was computed analytically for optimized entanglement swapping strategies, in dependence on four error parameters and efficiency, describing the (both initial and final) dephasing rate of the quantum memories, imperfect Bell state preparations and Bell state measurements. Replacing each quantum memory with an encoded version, we aim to correct errors due to imperfect initial Bell states, faulty Bell state measurements, and memory dephasing. Some of the more technical details of the approach of Ref. [15] are presented in Appendix A.

Extending the secret key rate analysis [15] to memory-corrected quantum repeaters involves determining effective error parameters for the logical qubit when its physical constituents are subject to noise. These logical error parameters are then substituted into the secret key rate formulas of Ref. [15]. To accomplish this, we introduce a novel method, inspired by the framework presented in Ref. [33], whereby we, however, eliminate the explicit usage of density matrices. We refer to this approach as the check matrix model and explain it in Sec. III B. Subsequently, we will apply this model to explore the behavior of stabilizer codes under the influence of noise. Lastly, we examine the performance of quantum repeaters with encoded quantum memories.

### A. Secret key rate and physical noise model

The secret key rate is the product of the secret key fraction, which characterizes the quality of the distributed state used to extract a secret key, and the raw rate, which is the speed at which the entangled state is distributed between the two end-node users. In the following, we always consider the protocol, where entanglement is distributed in parallel and swapped as soon as possible, which we will refer to as the optimal scheme [15].

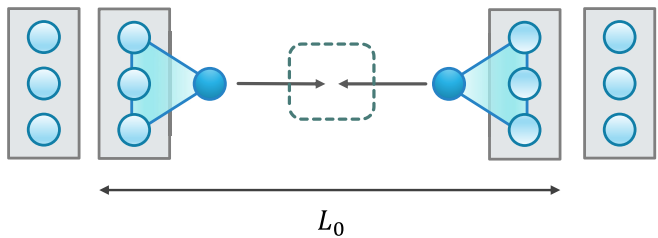

FIG. 1. For encoded entanglement distribution over one segment of length $L_0$, we first create a logical state between the memory qubits (light blue circles) and a photonic qubit. The photons (dark blue circles) at the end nodes of a segment are sent to the middle, where a photonic Bell state measurement is performed (dashed box). To connect neighboring segments, a logical Bell state measurement on the two quantum memories at the repeater node between them is performed (a joint measurement on the memory qubits in the two grey boxes that are present at each intermediate repeater node).

The raw rate depends on the success probability of one entanglement distribution attempt,

$$p(L_0) = p_0 e^{-\frac{L_0}{L_{\text{att}}}} = p_0 e^{-\frac{L_0}{22 \text{ km}}},$$

where $L_0$ is the elementary segment length of the quantum repeater. A typical attenuation distance due to transmission losses is $L_{\text{att}} = 22$ km (ultralow loss fibers can achieve attenuation distances between approximately 25 and 30 km) [15,36], and various efficiencies of the experimental hardware are included in a zero-distance link-coupling efficiency $p_0 = p(0)$, such as fiber coupling, wavelength conversion, or detector efficiencies.

For encoded entanglement distribution, we will require entanglement between the photonic state, our flying qubit, and the logical state of a quantum memory, our stationary qubit, so

$$\frac{1}{\sqrt{2}}(|0\rangle|0_L\rangle + |1\rangle|1_L\rangle). \tag{1}$$

A photonic Bell state measurement in the middle of each segment then allows to entangle two distant quantum memories, as shown in Fig. 1, and logical Bell state measurements on the two logical memory qubits in an intermediate station extend the entanglement. Note that for the first quantum error correction steps to work reliably, we have to assume that the locally generated, initial code states are free of any errors that may propagate through the encoding circuits, thus affecting too many physical qubits. In other words, we assume that the initial, locally encoded states in Eq. (1) can be prepared in a fault-tolerant fashion [17,37,38].

To determine the secret key fraction, we require the noisy state shared at the end nodes of the quantum repeater. Here we shall introduce the noise model for the physical qubits. The errors introduced on a quantum state $\rho$ due to memory storage, while waiting for a neighboring segment to become ready for entanglement swapping, are modelled via a dephasing channel,

$$\mathcal{M}_{\lambda_Z}(\rho) = (1 - \lambda_Z)\rho + \lambda_Z Z\rho Z.$$

Assuming that $0 \leqslant \lambda_Z < 1/2$, one can rewrite

$$\lambda_Z = (1 - e^{-\alpha})/2, \tag{2}$$

for some $\alpha > 0$ such that the dephasing channel becomes

$$\mathcal{M}_\alpha(\rho) = \frac{1 + e^{-\alpha}}{2}\rho + \frac{1 - e^{-\alpha}}{2}Z\rho Z. \tag{3}$$

We assume deterministic entanglement swapping, which is reasonable from an experimental point of view, given deterministic quantum gate implementations on stationary qubits and greatly simplifying rate optimizations. Probabilistic entanglement swapping based on heralded but nondeterministic Bell state measurements requires a more complicated treatment [39,40]. The Bell state measurement errors are described via two-qubit depolarization,

$$\mathcal{B}_\mu = \mu\rho + (1 - \mu)\frac{I}{4}.$$

Hence, the imperfect Bell state measurement is modelled by applying this depolarizing channel, followed by the ideal, noiseless Bell state projection.

The initial states are then modelled as imperfect Bell states with some initial depolarization $\mu_0$,

$$\rho_0 = \mathcal{B}_{\mu_0}(|\Psi^+\rangle\langle\Psi^+|),$$

where $|\Psi^+\rangle = \frac{1}{\sqrt{2}}(|01\rangle + |10\rangle)$ is a particular Bell state. The calculation of the final shared noisy state $\rho_N$ for an $N$-segment repeater and from that, the secret key rate, is detailed in Appendix A. Notably, the quantum bit error rates that determine the secret key fraction depend on a factor $\mu_0^N\mu^{N-1}$ meaning that scaling up the quantum repeater to higher segment numbers $N$ strongly impacts the secret key fraction for nonunit $\mu$ and $\mu_0$. This leads to high experimental requirements on the minimal depolarization parameters $\mu$ and $\mu_0$ for a nonzero secret key rate. For simplicity, we shall later assume $\mu_0 = \mu$.

The parameter $\alpha$ turns out to be not a mere mathematical convenience but has a physical meaning as the effective inverse coherence time,

$$\alpha(L_0) = \frac{\tau}{t_c}, \tag{4}$$

where $t_c$ is the coherence time of the quantum memory and $\tau$ is the elementary time unit of the quantum repeater. Since one entangled pair in two distant quantum memories dephases during each time step, we will double $\alpha$ for a given coherence time $t_c$ in the secret key rate analysis. For segment lengths of down to 10 km, the local processing time is negligible compared to the transmission time $L_0/c_f$ with $c_f$ being the speed of light in the optical fiber assuming an index of refraction of $n_r = 1.44$, whereas for smaller distances the repetition rate $\tau_{\text{clock}}$ becomes increasingly relevant. Therefore one needs to include the clock rate in the elementary time unit of the quantum repeater,

$$\tau = \tau_{\text{clock}} + \frac{L_0}{c_f}. \tag{5}$$

We assume typical experimental repetition rates of $\tau_{\text{clock}}^{-1} = 1$ MHz [6,15]. Note that a point-to-point optical link can usually operate at GHz clock rates since neither light-matter coupling of MHz order nor additional classical communication for confirming the successful transfer of entangled photons are required, as it is the case for memory-assisted quantum communication.

### B. The check matrix model

To extend the secret key rate analysis to memory-corrected quantum repeaters, we shall now introduce the check matrix model. Previous work [32] has described how to efficiently incorporate Pauli noise processes into the stabilizer formalism whenever the final state size is small, for instance, if a potentially large initial state is manipulated with local measurements. However, for our purposes, the number of qubits will remain unaltered. Mixed stabilizer states [31,41] describe a subclass of mixed states where the stabilizer group of $n$ qubits has $k < n$ independent generators. However, applying noise on stabilizer states does not in general result in mixed stabilizer states.

The check matrix model is inspired by the idea of translating physical single-qubit Pauli noise channels into an effective mean Pauli noise channel, acting on a logical qubit [33]. In contrast to Ref. [33], we avoid calculations in the state picture, allowing us to go beyond the five-qubit code up to a stabilizer code of eleven qubits. Moreover, we easily generalize the model to two-qubit Pauli noise channels.

Our approach is based on associating every possible multiqubit Pauli error after correction with either the logical identity $I_L$ or a logical Pauli error $X_L, Y_L$, or $Z_L$. This is possible since it is already established in Ref. [33] that physical Pauli noise channels acting on stabilizer codes result in a mean logical Pauli noise channel (see Appendix B). Furthermore, for Pauli-diagonal quantum noise channels, global phases of Pauli group elements cancel and it is sufficient to consider their row vector representation. All calculations are, thus, done using the check matrix of a given stabilizer code of $n$ qubits and the corresponding binary representation $r(g)$ of a Pauli element $g \in G_n$.

First, we apply single-qubit Pauli noise channels

$$\mathcal{E}(\rho) = \lambda_0 \rho + \lambda_1 X \rho X + \lambda_2 Y \rho Y + \lambda_3 Z \rho Z \qquad (6)$$

to every physical qubit composing a logical qubit of a stabilizer code. Subsequently, error correction is performed, which yields a Pauli noise channel on the logical level

$$\mathcal{E}_L(\rho_L) = \lambda_I \rho_L + \lambda_X X_L \rho_L X_L + \lambda_Y Y_L \rho_L Y_L + \lambda_Z Z_L \rho_L Z_L, \qquad (7)$$

whereby $\lambda_I$ is the probability of successful recovery, and $\lambda_X$, $\lambda_Y$, $\lambda_Z$ are the mean error probabilities for a logical Pauli error $X_L, Y_L$, or $Z_L$, respectively, to occur. The original error parameters $\lambda_1, \lambda_2$, and $\lambda_3$ will be parameterized in terms of an error probability $p$ that characterizes the respective quantum noise channel. For instance, a depolarizing channel with depolarization probability $p$ that transforms a single-qubit density matrix $\rho$ according to

$$\rho \longrightarrow (1-p)\rho + p\frac{I}{2}$$

is a Pauli noise channel, setting $\lambda_1 = \lambda_2 = \lambda_3 = \frac{p}{4}$ and $\lambda_0 = 1 - \frac{3}{4}p$ in Eq. (6), which is essentially the operator-sum representation of a depolarizing channel [5]. A dephasing channel has $\lambda_1 = \lambda_2 = 0$ and $\lambda_3 = p$ with $p$ being now the phase flip probability.

The check matrix model is easily generalized to two-qubit Pauli noise channels

$$\mathcal{E}(\rho) = \lambda_{00} \rho + \lambda_{01}(I \otimes X)\rho(I \otimes X)$$
$$+ \cdots + \lambda_{33}(Z \otimes Z)\rho(Z \otimes Z),$$

which includes two-qubit depolarization, described by the quantum channel

$$\rho \longrightarrow (1-p)\rho + p\frac{I}{4},$$

whereby all fifteen two-qubit Pauli errors are equiprobable, namely, $\lambda_{ij} = \frac{p}{16}$ for all $(i, j) \in \{0, 1, 2, 3\}^2 \backslash (0, 0)$. This channel is also translated into a logical two-qubit Pauli noise channel that has fifteen effective error parameters $\lambda_{AB}$ with $(A, B) \in \{I, X, Y, Z\}^2 \backslash (I, I)$. Since renaming the logical qubits does not affect what physically happens, we always have $\lambda_{AB} = \lambda_{BA}$.

The first step of the check matrix model is to partition the physical state space, similar to Ref. [33], into two-dimensional subspaces, namely, the code space $V_0$ and $2^{n-1} - 1$ error spaces $V_i$ as shown in Fig. 2. The error spaces are constructed by iterating over either general Pauli errors or dephasing errors, depending on how we choose the error syndrome identification. This choice is the essence of our adaptive error syndrome identification strategy. More precisely, besides correcting more general Pauli errors from the imperfect initial states and the imperfect swapping gates, if we want to specifically correct the effect of a memory dephasing channel, we check whether phase flip error syndromes describe all possible distinct error spaces. If the answer is positive, our error syndrome identification is tailored to the physical noise that the logical states experience. Furthermore, note that in general, it is only necessary to identify the possible syndromes with errors, so to partition only the state space that our logical subspace can be transported to by the Pauli noise channel instead of the entire physical state space. We start with single-qubit errors, continuing with two-qubit errors, and so on until all possible syndromes have been mapped to errors.

For every error $E \in G_n$ under consideration, its syndrome is computed by multiplication with the check matrix, $f(E)^T = \mathcal{C}\Omega r(E)^T$, and then it is determined whether this syndrome is already associated with an error. If not, the error and syndrome are saved. These proceedings associate the minimal-weight error with a given syndrome. Additionally, since the error spaces correspond to different syndromes, states in different error spaces are orthogonal. To demonstrate this, suppose $|v\rangle \in V_i$ and $|w\rangle \in V_j$ with $i \neq j$, so there exists a generator $g_k$ which has different commutation relations with the errors $E_i$ and $E_j$, associated with the error spaces $V_i$ and $V_j$. Let us say that $g_k$ anticommutes with $E_i$ and commutes with

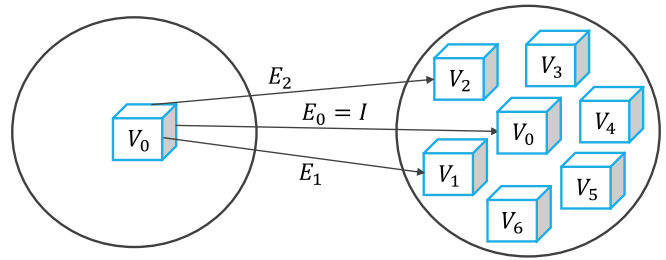

FIG. 2. State space partitioning. The physical state space of $n$ qubits is divided into the two-dimensional code space $V_0$ and two-dimensional orthogonal error spaces $V_i$, associated with errors $E_i \in G_n$ acting on $V_0$.

$E_j$. We then have

$$\langle v|w\rangle = \langle v|g_k|w\rangle = \langle v|g_k^\dagger|w\rangle = -\langle v|w\rangle = 0,$$

so the error spaces $V_i$ and $V_j$ are indeed orthogonal.

For every possible error $E \in \{I, X, Y, Z\}^{\otimes n}$, there is a certain occurrence probability $P(E)$ that depends on the Pauli noise channel. The calculation of $P(E)$ is detailed in Appendix C. We determine the syndrome $f(E)$, that each error causes, and the Pauli error $E_i$ that we have associated with that syndrome, so which satisfies $f(E) = f(E_i)$. Correction is then achieved by applying $E_i$ since Pauli errors are self-inverse. Thus, in the binary row vector representation, $r(E_i) + r(E)$ modulo two is computed. Afterwards, the total operation is equal to a logical Pauli operation $\sigma(E)$ (including the logical identity) up to stabilizers. To identify which logical operation was effectively performed on the encoded system we determine for which $\sigma_L \in \{I_L, X_L, Y_L, Z_L\}$ it is $r(E) + r(E_i) + r(\sigma_L) \in r(S)$, setting $\sigma(E) \equiv \sigma_L$. Summing up all probabilities contributing to $I_L$, $X_L$, $Y_L$, and $Z_L$, we find a mean single-qubit Pauli noise channel as in Eq. (7):

$$\mathcal{E}_L(\rho_L) = \sum_E P(E)\sigma(E)\rho_L\sigma(E).$$

The stabilizer group $S$ is computed by taking the generators of the stabilizer group $g_1, \ldots, g_l$, where $l = n - 1$ for one logical qubit, and calculating the stabilizer group element $g_1^{x_1} \cdot \cdots \cdot g_l^{x_l}$ for every binary vector $x = (x_1, \ldots, x_l)$, so computing modulo two

$$x_1 r(g_1) + \cdots + x_l r(g_l).$$

For two-qubit Pauli noise channels, the stabilizer group and the error spaces are constructed in the same manner as for single-qubit Pauli noise channels. However, now, one deals with two encoded logical systems that experience two-qubit noise transversally on their respective physical qubits. Syndrome detection and error correction are performed independently on both logical qubits. We again associate each error $E$ on an encoded system to a logical operator $\sigma(E) \in \{I_L, X_L, Y_L, Z_L\}$. In this way, we obtain a logical two-qubit Pauli noise channel $\mathcal{E}_L(\rho_L)$ that has, in the most general case, sixteen effective parameters, one describing successful recovery and fifteen describing logical errors,

$$\sum_{E_1, E_2} P(E_1, E_2)(\sigma(E_1) \otimes \sigma(E_2))\rho_L(\sigma(E_1) \otimes \sigma(E_2)),$$

where $P(E_1, E_2)$ is the occurrence probability for the combined errors $E_1$ and $E_2$. The calculation of $P(E_1, E_2)$ for two-qubit depolarization is given in Appendix C.

The most inefficient part of the check matrix model is that the number of possible errors increases exponentially with $2^{2n} = 4^n$ for general Pauli noise channels. Also, the stabilizer group size is given by $2^{n-1}$. Nonetheless, the method proved to be much more efficient compared to the effective Pauli noise channel calculation presented in Ref. [33], which requires working with $2^n \times 2^n$ matrices to apply noise in the state picture, to perform error detection with $2^{n-1}$ error space projectors, and to then error-correct. Extra remarks concerning the algorithmic implementation of the check matrix model for runtime speed-up can be found in Appendix C. The explicit analytical expressions for the logical error parameters are given in Appendix D.

The model from Ref. [33] and the check matrix model were found to yield the same mean error probabilities in the case of single-qubit depolarization for all considered codes apart from the eleven-qubit code, which was too runtime-inefficient to calculate with the former model. However, using solely the methods from Ref. [33], results could only be obtained when sampling the probability values starting from the seven-qubit Steane code already—the runtime cost of symbolic computations for large density matrices quickly became too high—whereas, with the check matrix model, we obtained exact analytical expressions up to the eleven-qubit code.

## IV. RESULTS

Our main interest is to understand which quantum stabilizer codes increase resilience against depolarizing and dephasing noise and in what experimental parameter regime the application of encoded quantum memories is of advantage. Therefore, in Sec. IV A, we investigate for several stabilizer codes how error probabilities transform when we move from physical noise channels to logical noise channels.

After establishing the five-qubit and the Steane code as particularly promising quantum memory encodings, we study memory-corrected quantum repeaters in Sec. IV B by applying the effective error parameters from Sec. IV A to the secret key rate analysis [15].

### A. Effective logical noise channels

#### 1. The three-qubit bit flip and phase flip codes

A simple idea to protect quantum information against bit flip errors is a repetition code [5], where one defines the logical zero state as $|0_L\rangle \equiv |000\rangle$ and the logical one via $|1_L\rangle \equiv |111\rangle$. The states $|000\rangle$ and $|111\rangle$ are stabilized by $Z_1Z_2$ and $Z_2Z_3$, which are the generators of the three-element stabilizer group $S = \{I, Z_1Z_2, Z_2Z_3, Z_1Z_3\} = \langle Z_1Z_2, Z_2Z_3\rangle$. The three-qubit repetition code can correct bit flips on up to one qubit and its logical operators are given by $Z_L = Z_1Z_2Z_3$ and $X_L = X_1X_2X_3$. Here and throughout, our particular choice of logical operators is to be understood, as usual, up to multiplication with stabilizer group elements. In a similar manner, we can protect a logical qubit against phase flips using $|0_L\rangle = |+++\rangle$ and $|1_L\rangle = |---\rangle$, so working in the single-qubit basis $|+\rangle = \frac{1}{\sqrt{2}}(|0\rangle + |1\rangle)$ and $|-\rangle = \frac{1}{\sqrt{2}}(|0\rangle - |1\rangle)$. In the $\{|+\rangle, |-\rangle\}$ basis, the phase flip operator $Z$ takes $|+\rangle$ to $|-\rangle$ and vice versa so it acts just like the bit flip operator $X$. Thus the stabilizer generators become $X_1X_2, X_2X_3$ and the logical operators are $Z_L = X_1X_2X_3, X_L = Z_1Z_2Z_3$.

Consider a dephasing channel acting on the three-qubit bit-flip and phase-flip codes. The resulting logical channels are shown in Fig. 3(a) and the new phase flip probability for the case of the bit-flip code is given by

$$\lambda_Z = p^3 + 3p(1 - p)^2,$$

which is simply the probability that either one or three qubits experience a phase flip. This can be attributed to phase flips on two qubits leaving a bit-flip code of three qubits invariant. Even though the three-qubit bit flip code has no means of detecting phase flips, this result shows that the logical error probabilities not only manifest themselves when error detection or correction is performed but are also determined by the encoding itself. The three-qubit phase flip code can correct one phase flip, however, two phase flips or more manifest themselves as a bit flip on the logical level with

$$\lambda_X = p^3 + 3p^2(1-p),$$

as can be seen in Fig. 3(a). However, a phase flip channel has, per definition, zero $X$ and $Y$ error probabilities, and setting nonzero logical error probabilities to zero would improve the quantum noise channel, so that we would only be able to compute upper bounds for the secret key rate which do not allow to make any meaningful statements.

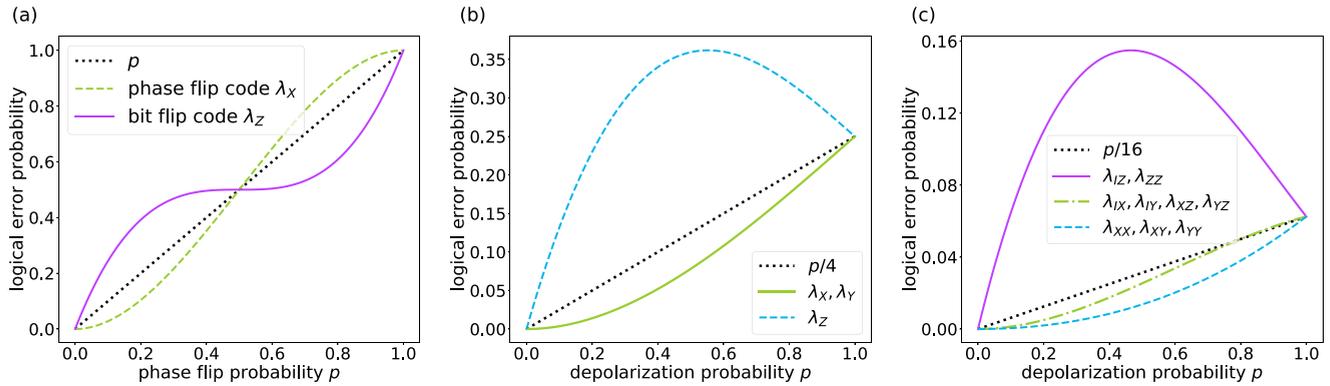

FIG. 3. Logical error parameters for the three-qubit repetition bit and phase flip codes subject to (a) single-qubit dephasing, (b) single-qubit depolarization, and (c) two-qubit depolarization.

The logical error probabilities, associated with corrected single-qubit depolarization, can be seen in Fig. 3(b) and are equal for the three-qubit bit flip and phase flip repetition codes. For comparison, the original error probability $\lambda_1 = \lambda_2 = \lambda_3 = p/4$ in the operator-sum representation of a depolarizing channel is depicted. The mean error probability $\lambda_Z$ is enhanced since the bit flip code is unable to protect a logical qubit against phase flips, whereas $X$ and $Y$ errors are equally reduced, $\lambda_X = \lambda_Y$. For the phase flip code, it seems counterintuitive that logical phase flips are enhanced, however, this can be attributed to the logical $Z_L$ operator consisting of physical bit flips $X$, so the roles of $X_L$ and $Z_L$ are swapped compared to the bit flip code.

When the three-qubit repetition codes are subject to two-qubit depolarization, see Fig. 3(c), the resulting logical channel is again equal for the bit flip and phase flip codes. The errors that increase the effective error probabilities compared to the unencoded situation are those involving phase flips and no bit flips. What can be treated the best are $X_L$ or $Y_L$ errors on both logical qubits which reflects the observed behavior against single-qubit depolarization. Although most of the error probabilities have improved, we cannot represent the effective two-qubit logical channel in terms of an improved depolarizing channel due to the increased phase flip error rates. In our later worst-case treatment, we approximate directed logical two-qubit Pauli noise, which can have fifteen distinct error parameters in the most general case, with an undirected two-qubit depolarizing channel as in Ref. [15], described via only one error parameter, namely, the two-qubit depolarization probability. The logical two-qubit depolariza-tion probability is later computed from the highest, so worst, logical error parameter. Thus the secret key rate only improves if even the worst logical error parameter is below the physical error probability, which does not hold for the three-qubit bit and phase flip codes. The latter would reduce the effective memory errors, however, resulting in an effective logical bit flip channel that can currently not be incorporated into the exact secret rate analysis of Ref. [15] that does not include time-dependent, statistical errors other than random phase flips from physical memory dephasing. An extension of the analytical framework of Ref. [15] that allows for more general time-dependent, storage-induced errors and two-qubit Pauli noise channels instead of two-qubit depolarization could also be considered but is not the focus of the present work.

In total, we cannot report a simultaneous improvement of all logical error probabilities of the considered quantum noise channels and, lacking an extended framework of Ref. [15], we cannot analyze whether repetition codes improve the secret key rate nonetheless.

### 2. The five-qubit code

The five-qubit code is the smallest code that can correct arbitrary single-qubit errors. Its stabilizer group generators are given by

$$g_1 = X_1 Z_2 Z_3 X_4, \quad g_3 = X_1 X_3 Z_4 Z_5,$$
$$g_2 = X_2 Z_3 Z_4 X_5, \quad g_4 = Z_1 X_2 X_4 Z_5,$$

and the logical operators are $Z_L = Z_1 Z_2 Z_3 Z_4 Z_5$, $X_L = X_1 X_2 X_3 X_4 X_5$. Note that this code was recently implemented on nuclear spin qubits in diamond [42].

Since all fifteen single-qubit Pauli errors on a five-qubit code logical qubit are distinguishable, all fifteen error spaces could be constructed by considering all possible single-qubit Pauli errors. For single-qubit dephasing, we have adjusted the error space construction and error syndrome identification. Otherwise, without this adaptation, the resulting logical channel would have nonzero $\lambda_X$ and $\lambda_Y$ components, as displayed in Fig. 17(a) in Appendix E, thus becoming inapplicable to the secret key rate analysis of Ref. [15] that requires a pure dephasing channel. Because we aim at detecting phase flips, we now use minimal-weight multiqubit phase flip errors to search for different error syndromes. It turned out to be sufficient to consider only single and two-qubit phase flips to find fifteen distinct syndromes and, thus, error spaces. This means that all five single-qubit phase flip error syndromes and all $\binom{5}{2} = 10$ two-qubit phase flip error syndromes are distinguishable. Therefore, with our adaptive error syndrome identification, the five-qubit code is able to successfully correct phase flips on up to two qubits, while, with the standard error syndrome identification, it is additionally able to correct arbitrary single-qubit errors. The logical phase flip probability is then calculated to be

$$\lambda_Z = p^5 + 5p^4(1-p) + 10p^3(1-p)^2,$$

which is just the probability that more than two phase flips occur on the physical level. It is plotted in Fig. 4(a). For a phase flip probability of 1%, it is apparent in the inset figure that the error probability improves by about three orders of magnitude to 0.001% with encoding. Interestingly, the result of applying single-qubit depolarization results also in a logical depolarizing channel which was already observed in Ref. [33] and can be seen in Fig. 4(b), where $\lambda_X = \lambda_Y = \lambda_Z$. In the inset figure with logarithmic effective error probabilities, it is evident that a depolarizing probability of around 1% can be improved by about one order of magnitude.

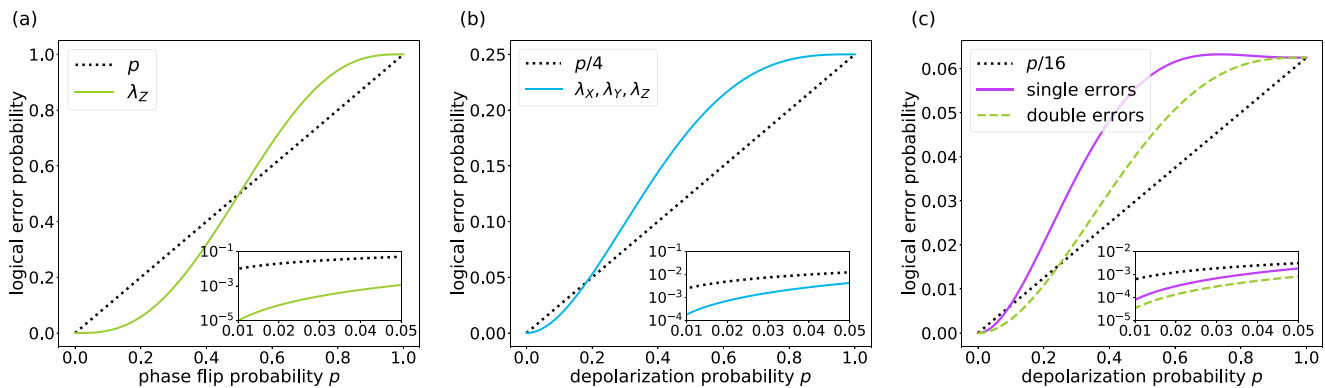

FIG. 4. Logical error parameters for the five-qubit code subject to (a) single-qubit dephasing, (b) single-qubit depolarization, and (c) two-qubit depolarization.

For a two-qubit depolarizing channel, we obtain two distinct error probability curves, as depicted in Fig. 4(c). The upper curve corresponds to logical single-qubit errors, namely, $\lambda_{IX} = \lambda_{XI}, \lambda_{IY} = \lambda_{YI}, \lambda_{IZ} = \lambda_{ZI}$, and the lower curve to logical errors on both qubits, so $\lambda_{XX}, \lambda_{YY}, \lambda_{ZZ}, \lambda_{XY} = \lambda_{YX}, \lambda_{XZ} = \lambda_{ZX}, \lambda_{YZ} = \lambda_{ZY}$. In the regime where errors are unlikely to occur, both curves stay below $p/16$ and so the encoding provides a benefit.

Overall, there is an advantage compared to the unencoded quantum states when using the five-qubit code to protect a logical qubit against depolarizing noise and particularly to protect it against dephasing noise. Due to our adapted error syndrome identification, a phase flip probability of 1% improves by a factor of around one thousand. Note that the ability to correct up to two random phase flips on five physical qubits, as exploited in our adaptive syndrome identification scheme for memory dephasing, would also be provided by a simple five-qubit phase flip repetition code with the same amount of physical resources. However, as already mentioned at the end of the preceding subsection, the effective logical channel for the five-qubit repetition code would be incompatible with the current memory dephasing model of Ref. [15]. Moreover, we would sacrifice the code's ability to additionally correct arbitrary Pauli errors as required in a complete encoded quantum repeater protocol.

### 3. The Steane code

The seven-qubit Steane code belongs to the class of Calderbank-Shor-Steane (CSS) codes [5], which are stabilizer codes based on the idea of using two classical linear codes, one to protect against bit flips and the other one against phase flips. The six generators of the Steane code,

$$g_1 = X_4X_5X_6X_7, \quad g_4 = Z_4Z_5Z_6Z_7,$$
$$g_2 = X_2X_3X_6X_7, \quad g_5 = Z_2Z_3Z_6Z_7,$$
$$g_3 = X_1X_3X_5X_7, \quad g_6 = Z_1Z_3Z_5Z_7$$

are illustrated in Fig. 5. Its logical operators are $Z_L = Z_1Z_2Z_3Z_4Z_5Z_6Z_7$ and $X_L = X_1X_2X_3X_4X_5X_6X_7$.

For our purposes, the error spaces of a Steane code logical qubit are constructed with multiqubit Pauli errors and it is sufficient to consider errors of up to weight two. Phase flip errors on at least two qubits now result in the same error syndromes as single-qubit phase flip errors. Thus there are no multiqubit phase flips that could be unambiguously identified in addition to single-qubit phase flips.

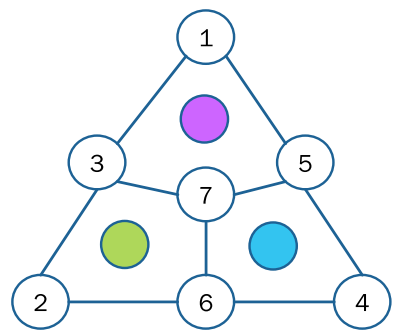

FIG. 5. Stabilizer generators of the Steane code as an instance of color codes [43,44], a subset of CSS codes. The physical qubits correspond to vertices and $X$- and $Z$-type stabilizer generators are associated with the violet, green, and blue colored faces.

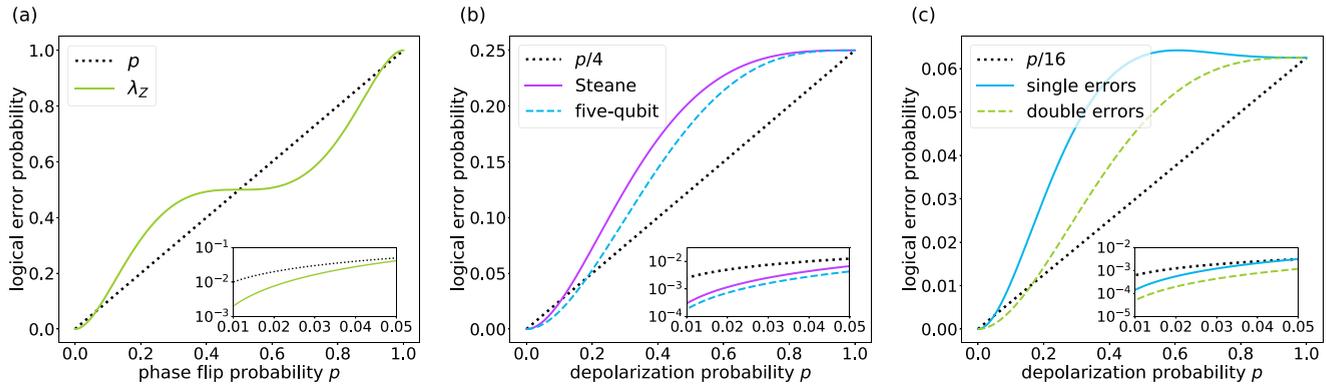

FIG. 6. Logical error parameters for the Steane code subject to (a) single-qubit dephasing, (b) single-qubit depolarization, and (c) two-qubit depolarization.

When the Steane code is subject to single-qubit dephasing, the result is a logical phase flip channel, see Fig. 6(a). Below $p \approx 0.05$, the phase flip probability is decreased compared to the original one, and a phase flip probability of 1% is reduced by around a factor of ten on the logical level. This improvement is less spectacular than for the five-qubit code, as there our error syndrome identification was particularly well adjusted to the modelled noise since we had picked multiqubit phase flips to construct the error spaces and could distinguish phase flips on up to two qubits.

In Fig. 6(b), one can see that applying single-qubit depolarization leads to a logical depolarizing channel. We observe that the logical error parameter $\lambda_X = \lambda_Y = \lambda_Z$ is higher for the Steane code than for the five-qubit code for all depolarization probabilities $p$. From a physical point of view, it is reasonable that one ends up with a depolarizing channel on the logical level when using the Steane code. This can be well understood by recalling that the six generators of the Steane code stabilizer group are three Pauli elements involving only $Z$ gates and three Pauli elements that can be obtained from these previous ones by simply replacing the $Z$ gates with $X$ gates. Since also operators obtained by multiplying generators are contained in the stabilizer group, we have three analogous Pauli elements involving only $Y$ gates stabilizing the code space. In total, there is no bias with regard to the error-correction ability, so we should have equal logical error probabilities. Consequently, we expect to find a depolarizing channel on the logical level for all codes that have analogous $X$ and $Z$ stabilizer group generators. One example is color codes, proposed in Ref. [43], which are CSS codes, with the seven-qubit Steane code being the smallest representative of this class of codes [44].

The effective error parameters after correcting two-qubit depolarization are displayed in Fig. 6(c). As has been already observed for the five-qubit code, we obtain two error curves corresponding to logical Pauli errors on a single qubit and on two qubits. For low depolarization probability $p$, starting from around $p \approx 0.05$, one has a regime, where the encoding provides an advantage as can be seen in the inset figure. The Steane code performs worse than the five-qubit code in every parameter regime both for single-qubit dephasing and depolarization as well as for two-qubit depolarization. A possible explanation is that one introduces more physical qubits that can be affected by noise without having a significant counterweight in terms of error-correction capabilities.

### 4. The nine-qubit surface code

Toric and surface codes were introduced by Kitaev [45]. They are a particular class of stabilizer codes defined on a torus or lattice, respectively. The generators of the nine-qubit surface code can be extracted from Fig. 7 as

$$
\begin{aligned}
g_1 &= X_1 X_2 X_4 X_5, & g_5 &= Z_1 Z_4, \\
g_2 &= Z_2 Z_3 Z_5 Z_6, & g_6 &= X_2 X_3, \\
g_3 &= Z_4 Z_5 Z_7 Z_8, & g_7 &= X_7 X_8, \\
g_4 &= X_5 X_6 X_8 X_9, & g_8 &= Z_6 Z_9.
\end{aligned}
$$

The logical operators are given by $Z_L = Z_1 Z_2 Z_3$ and $X_L = X_1 X_4 X_7$. All errors that do not connect two opposite boundaries of the lattice can be corrected, so the distance of the nine-qubit surface code is three and it can correct arbitrary errors on one physical qubit. One motivation for considering this surface code lies in the resilience of topological codes against depolarizing noise [46]. Recently, there were also several advances in the experimental implementation of this code, in a superconducting circuit [47] and on a superconducting quantum processor [48].

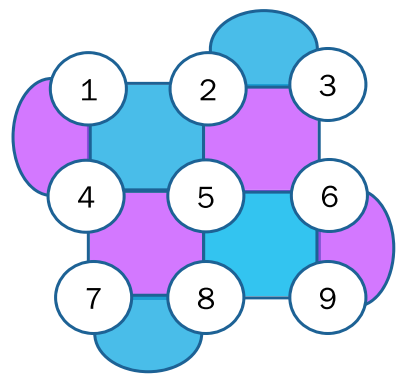

FIG. 7. Stabilizer generators of the nine-qubit surface code. The physical qubits correspond to vertices, $X$-type stabilizer group generators are associated with the blue faces, and $Z$-type generators with the violet faces.

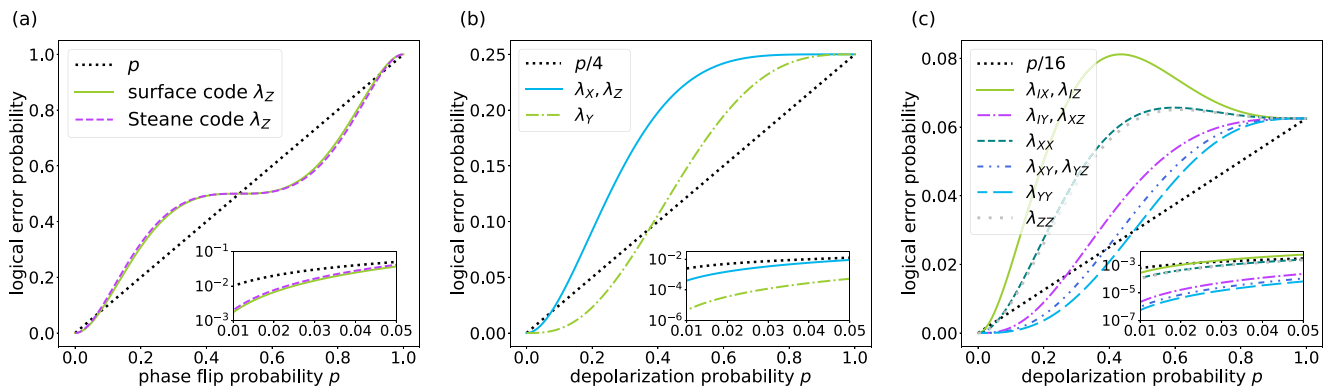

FIG. 8. Logical error parameters for the nine-qubit surface code subject to (a) single-qubit dephasing, (b) single-qubit depolarization, and (c) two-qubit depolarization.

For the surface code of nine qubits, it is necessary to consider multiqubit Pauli errors of up to weight four to span the entire physical space. When considering multiqubit phase flip errors, starting from errors on three qubits, no new error syndromes emerge. There are eight distinct two-qubit phase flip error syndromes which are, however, associated with the same phase flip errors irrespective of whether we perform minimal-weight Pauli error syndrome identification or minimal-weight phase flip error syndrome identification. Thus like for the Steane code, the adaptive error syndrome identification leads to the same effective quantum noise channel after single-qubit dephasing as does the standard syndrome identification.

Single-qubit phase flips yield a logical dephasing channel, as displayed in Fig. 8(a). The logical phase flip probability qualitatively coincides with that of the Steane code after correcting single-qubit dephasing; for low error probabilities, the nine-qubit surface code performs slightly better in terms of reducing the effective noise. Hence, the benefit of using a larger stabilizer code with better error-correction capabilities seems almost entirely outweighed by the disadvantage of having more qubits that can be affected by errors.

When applying single-qubit depolarization and after performing error correction, an effective mean logical channel is produced with almost equal $\lambda_X$ and $\lambda_Z$ error parameters, and a differing $\lambda_Y$ error parameter, see Fig. 8(b). Visually, $\lambda_X$ and $\lambda_Z$ overlap, however, interestingly, the analytical expressions are not exactly equal. Still, this result is concordant with intuition since the stabilizer group generators of the nine-qubit surface code are defined analogously for the four generators involving $X$ gates and the four generators involving $Z$ gates.

Corrected two-qubit depolarization, see Fig. 8(c), produces an effective mean logical channel with analytically nine different error parameters from which only six error curves can be visually distinguished since $\lambda_{IX} \approx \lambda_{IZ}$, $\lambda_{IY} \approx \lambda_{XZ}$, and $\lambda_{XY} \approx \lambda_{ZY}$. The lowest curve is associated with $Y$ errors on both logical qubits and all three lower error curves involve $Y$ errors (with $\lambda_{XZ}$ being an exception), whereas the three upper curves correspond to either $X$ or $Z$ errors. This is consistent with the logical channel obtained when the surface code is subject to single-qubit depolarization, displayed in Fig. 8(b), where the $\lambda_Y$ parameter is also lower than $\lambda_X \approx \lambda_Z$. Due to the $\lambda_{IX} \approx \lambda_{IZ}$ error parameter, the nine-qubit surface code provides a benefit against all possible Pauli errors starting only from depolarization probabilities below approximately 2.5% as can be seen in the inset figure.

Ultimately, we are interested in improving single-qubit dephasing and two-qubit depolarization. To model a logical two-qubit depolarizing channel, we pick the highest, so worst, error curve of the logical channel in Fig. 8(c) to get an upper bound on the error probabilities. The highest two-qubit error parameters, namely, $\lambda_{IX} \approx \lambda_{IZ}$, are well above the error probabilities of the Steane code, subject to two-qubit depolarization. Together with the logical phase flip probabilities being of a very similar magnitude to those for the Steane code, the nine-qubit surface code is expected to perform significantly worse than the Steane code. Therefore it will not be further investigated for the secret key rate analysis.

Another way to approximate a logical two-qubit depolarizing channel might be to average over all fifteen error parameters. In this case, the resulting mean error curve of the nine-qubit surface code is only slightly above the error probability for single-qubit logical errors of the five-qubit code in Fig. 4(c) for depolarization probabilities below 5%. So in terms of an averaged logical channel, the nine-qubit surface code performs almost as well as the five-qubit code. However, single-qubit dephasing is still corrected significantly better with the five-qubit code.

### 5. The Shor code

The Shor code [5] is based on the idea of concatenating the three-qubit bit flip and phase flip repetition codes to achieve protection against both. This yields a nine-qubit code with logical basis states as follows:

$$|0_L\rangle = \frac{(|000\rangle + |111\rangle)(|000\rangle + |111\rangle)(|000\rangle + |111\rangle)}{2\sqrt{2}},$$

$$|1_L\rangle = \frac{(|000\rangle - |111\rangle)(|000\rangle - |111\rangle)(|000\rangle - |111\rangle)}{2\sqrt{2}}.$$

The corresponding eight stabilizer group generators are

$$g_1 = Z_1 Z_2, \quad g_5 = Z_7 Z_8,$$
$$g_2 = Z_2 Z_3, \quad g_6 = Z_8 Z_9,$$
$$g_3 = Z_4 Z_5, \quad g_7 = X_1 X_2 X_3 X_4 X_5 X_6,$$
$$g_4 = Z_5 Z_6, \quad g_8 = X_4 X_5 X_6 X_7 X_8 X_9.$$

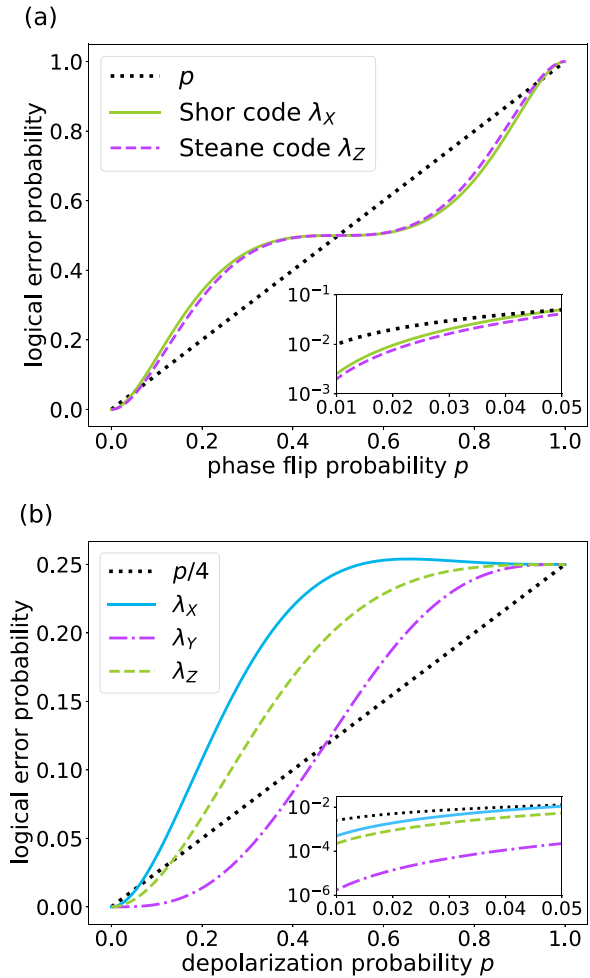

FIG. 9. Logical error parameters for the nine-qubit Shor code subject to (a) single-qubit dephasing and (b) single-qubit depolarization.

At first glance counter-intuitively, the logical operators of the Shor code are $Z_L = X_1X_2X_3X_4X_5X_6X_7X_8X_9$ and $X_L = Z_1Z_2Z_3Z_4Z_5Z_6Z_7Z_8Z_9$. However, considering the explicit logical basis states, it is easy to see that $Z_L|0_L\rangle = |0_L\rangle$, $Z_L|1_L\rangle = -|1_L\rangle$, $X_L|0_L\rangle = |1_L\rangle$, and $X_L|1_L\rangle = |0_L\rangle$.

For the Shor code, it is sufficient to consider up to three-qubit Pauli errors to span the whole physical space. We have only three nontrivial error syndromes due to phase flips since there are only two generators with which phase flips on multiple qubits could anticommute. Therefore the adaptive error syndrome identification cannot change the quantum noise channel after single-qubit dephasing.

When applying physical dephasing channels to a Shor code logical qubit, the result is a logical bit flip channel, see Fig. 9(a). Additionally, in Fig. 9(a), the Steane code logical phase flip probability $\lambda_Z$ is depicted because of the striking resemblance of the curve shape. While an effective logical bit flip channel acting on the Shor code may seem surprising, it is, in fact, rather simple to explain. Considering the computational basis states of the Shor code, applying $Z$ gates either does nothing or changes a block like $|000\rangle + |111\rangle$ to $|000\rangle - |111\rangle$ and vice versa. When majority voting for correcting phase flips on individual qubits is performed, either the state is recovered or the logical qubit flipped. Essentially this is the same explanation as to why $X_L = Z_1Z_2Z_3Z_4Z_5Z_6Z_7Z_8Z_9$. Since we have turned a dephasing channel into a bit flip channel on the logical level, the Shor code is not directly applicable to the secret key rate analysis of Ref. [15].

Single-qubit depolarizing channels do not yield an effective depolarizing channel on the logical level, as can be seen in Fig. 9(b). This can be attributed to the asymmetry in the stabilizer group generators of the Shor code, which involve only two generators with solely $X$ gates apart from identities, whereas six generators consist of only $Z$ gates. For single-qubit depolarization, the smallest improvement is associated with bit flips; at a depolarization probability of $p = 0.01$, the gain is less than a factor of ten which is worse than for the five-qubit code and similar to the Steane code.

### 6. The eleven-qubit code

The smallest code that can correct arbitrary errors on two qubits consists of eleven qubits. Its generators can be found to be [4]

$$
\begin{aligned}
&g_1 = Z_1Z_2Z_3Z_4Z_5Z_6, &&g_6 = X_1X_2X_3X_4X_5X_6, \\
&g_2 = Z_1Y_2X_3Z_7Y_8X_9, &&g_7 = Z_4X_5Y_6Y_7Y_8Y_9X_{10}Z_{11}, \\
&g_3 = X_1Z_2Y_3X_7Z_8Y_9, &&g_8 = X_4Y_5Z_6Z_7Z_8Z_9Y_{10}X_{11}, \\
&g_4 = Z_4Y_5X_6X_7Y_8Z_9, &&g_9 = Z_1X_2Y_3Z_7Z_8Z_9X_{10}Y_{11}, \\
&g_5 = X_4Z_5Y_6Z_7X_8Y_9, &&g_{10} = Y_1Z_2X_3Y_7Y_8Y_9Z_{10}X_{11}.
\end{aligned}
$$

Its logical operators may be expressed as $Z_L = Z_7Z_8Z_9Z_{10}Z_{11}$ and $X_L = X_7X_8X_9X_{10}X_{11}$.

With multiqubit Pauli errors, it is then sufficient to go up to weight-three errors. Our adaptive error syndrome identification allows to distinguish all eleven single-qubit phase flips (as does the minimal-weight Pauli error identification), 55 two-qubit phase flips (compared to 48 distinct error syndromes for nonadaptive error syndrome identification), 108 three-qubit phase flips (compared to 12), 66 four-qubit phase flips, and 18 five-qubit phase flips.

The logical quantum noise channel after single-qubit dephasing is then a phase flip channel, as shown in Fig. 10(a), and the eleven-qubit code performs almost as well as the five-qubit code. However, applying single-qubit dephasing without adaptive error syndrome identification yields a logical channel with nonzero $\lambda_X$ and $\lambda_Y$ components, see Fig. 17(b) in Appendix E. This can also be observed for the five-qubit code when constructing the error spaces with multiqubit Pauli errors.

Single-qubit depolarization acting on the eleven-qubit code does not yield a logical depolarizing channel, as shown in Fig. 10(b). The highest error probability is now associated with bit flips, whereas $\lambda_Y$ and $\lambda_Z$ seem to overlap even though the analytical expressions are not exactly equal, yet very similar.

The initial motivation was that the eleven-qubit code, being able to correct arbitrary errors on two qubits, might provide better resilience against dephasing or depolarizing noise than the five-qubit code despite the larger number of its physical constituents. Our result now is that with adaptive error syndrome identification, the code greatly reduces the effective dephasing, albeit performs slightly worse than the five-qubit code. The error parameter $\lambda_X$ after single-qubit depolarization is similar to the logical error parameter of the five-qubit code, whereas $\lambda_Y$ and $\lambda_Z$ are lower by about one order of magnitude, indicating that the performance of the eleven-qubit code against two-qubit depolarizing noise might be slightly better. Overall, we expect the eleven-qubit code to roughly match the five-qubit code in protecting a logical qubit against a combination of dephasing and depolarizing noise. However, the eleven-qubit code has the clear disadvantage of a higher resource overhead compared with the five-qubit code.

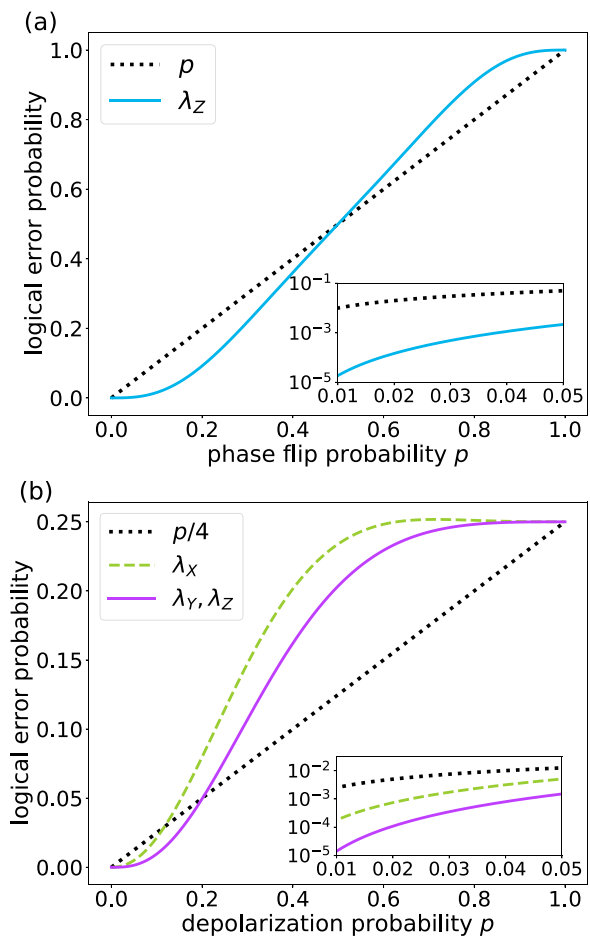

FIG. 10. Logical error parameters for the eleven-qubit code subject to (a) single-qubit dephasing, and (b) single-qubit depolarization.

## B. Secret key rates with encoding

For the secret key rate analysis, we consider the two most promising quantum memory encodings, namely, the five-qubit code and the Steane code. We assume that entanglement is distributed in parallel and swapped via a logical Bell state measurement as soon as possible, i.e., as soon as two adjacent segments each contain a successfully distributed entangled state. We refer to this as the optimal scheme since it reduces the memory dephasing as much as possible for a minimal total waiting time of the quantum repeater; as such, it maximizes the secret key rate in a three-segment repeater in the most relevant parameter regimes [15]. For a two-segment repeater, we additionally consider a memory cutoff where the storage of an entangled state is interrupted and entanglement distribution reinitiated when the total number of attempts in a neighboring segment has exceeded a prespecified value.

It is straightforward to see why encoding can reduce effective errors introduced through quantum memory dephasing, modelled via phase flip channels, and due to faulty Bell state preparation and measurements, modelled via two-qubit depolarization. We shall use the highest effective error probability curve of the logical two-qubit channel, obtained from physical two-qubit depolarization, to calculate a lower bound on the secret key rate. Thus to make sure that the effective channel remains depolarizing and is compatible with the exact rate analysis of Ref. [15], we approximate the resulting two-qubit logical Pauli channel with a two-qubit depolarizing channel by replacing all fifteen effective error parameters with the worst logical error parameter. Naturally, this worst-case approximation deteriorates the probability of successful error correction. Then, we can directly apply the effective error probability results to the secret key rate analysis of Ref. [15] and extend the model to memory-corrected quantum repeaters.

Herein, we mainly use the same experimental parameter values as in Ref. [15]. For instance, we assume there is no initial dephasing (though typically there always is due to the fixed waiting time during each individual, possibly even successful distribution attempt, but any such initial state imperfections are to be covered by a nonzero initial depolarization $\mu_0 < 1$), whereas we assume nonunit $\mu_0$ and $\mu$, meaning that we aim at correcting both Bell measurement induced depolarization and initial depolarization accounting for the imperfect preparation and distribution of the initially shared Bell states in each repeater segment. Furthermore, we set $\mu_0 = \mu$, so Bell measurement and Bell state preparation errors are assumed to be of equal size. This is a useful simplification [15], however, the encoded secret key rate analysis could also be performed in a more general setting, possibly better matching actual experiments with distinct error parameter values. In practice, we can correct initial depolarization directly after successfully distributing a state. To highlight the impact of the encoding we, however, take $\mu = 0.99$ instead of $\mu = 1$ as an improved Bell state measurement parameter. Taking the highest obtained error parameter $\lambda_{AB}$, $(A, B) \in \{I, X, Y, Z\}^2 \backslash (I, I)$, we have an upper bound on the factor $p/16$ of a logical two-qubit depolarizing channel, which we then translate into an effective success probability $\mu = 1 - p$.

By directly inserting effective error probabilities into the secret key rate formula, we implicitly assume that a logical Bell state measurement on the two logical quantum memories placed in one station is readily available and can be done transversally, so acting bitwise between corresponding qubits of each encoding block. This is rather straightforward for the Steane code, where both the Hadamard and the CNOT gates can be implemented transversally [49]. However, a Bell state measurement can, in fact, be implemented transversally for any stabilizer code [50]. To understand this, first note that entanglement swapping via a logical Bell state measurement is equivalent to measuring the observables $X_1^L X_2^L$ and $Z_1^L Z_2^L$ since Bell states are their unique eigenstates. Now $X_1^L X_2^L$ and $Z_1^L Z_2^L$ consist of the operators $X_1 X_2$, $Z_1 Z_2$, or $Y_1 Y_2$ acting on transversal pairs of qubits. Therefore, measuring the commuting, so compatible observables $X_1 X_2$ and $Z_1 Z_2$ for each pair allows to infer sufficient information to reconstruct the result of a measurement of both $X_1^L X_2^L$ and $Z_1^L Z_2^L$. Moreover, it might facilitate obtaining the error syndrome [51].

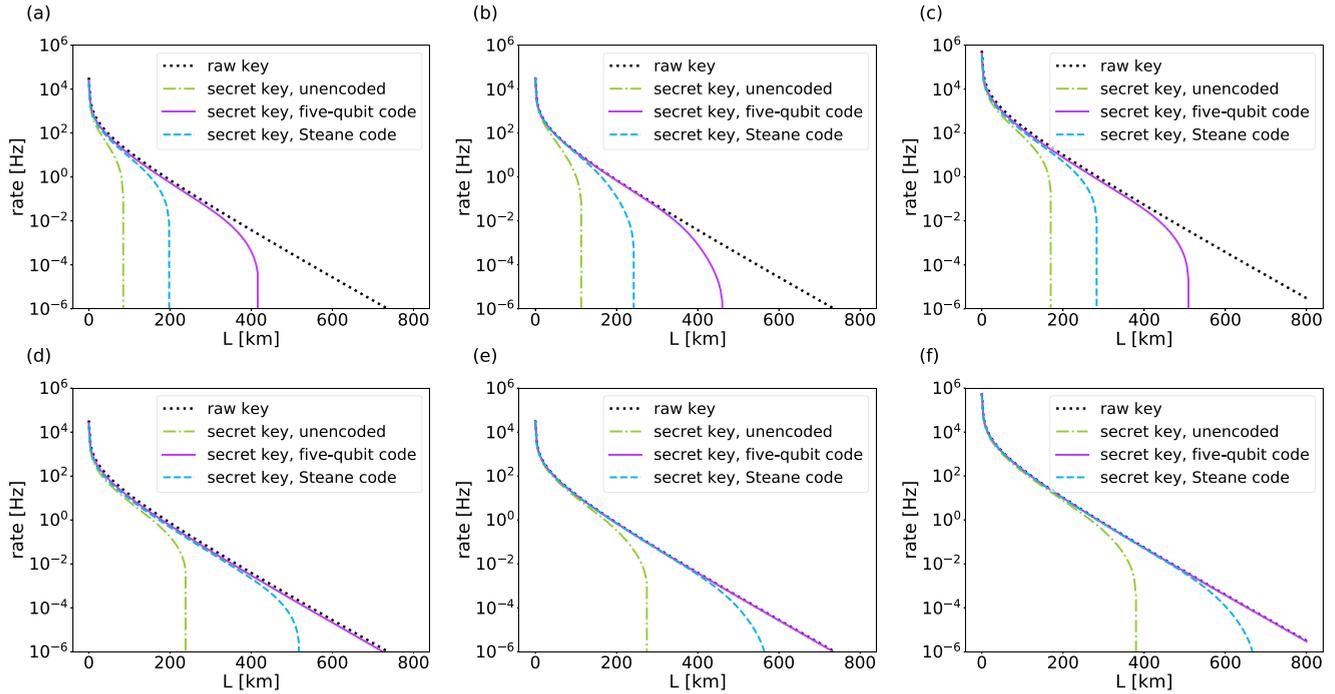

FIG. 11. Raw and secret key rates of a two-segment quantum repeater of length $L$ without memory cutoff for different quantum memory coherence times $t_c$, zero-distance link-coupling efficiencies $p_0$, Bell state depolarization parameters $\mu$, and initial depolarizations $\mu_0 = \mu$. The subplots correspond to the experimental parameters: (a) $t_c = 0.1$ s, $p_0 = 0.05$, $\mu = 0.97$; (b) $t_c = 0.1$ s, $p_0 = 0.05$, $\mu = 0.99$; (c) $t_c = 0.1$ s, $p_0 = 0.7$, $\mu = 0.97$; (d) $t_c = 10$ s, $p_0 = 0.05$, $\mu = 0.97$; (e) $t_c = 10$ s, $p_0 = 0.05$, $\mu = 0.99$; and (f) $t_c = 10$ s, $p_0 = 0.7$, $\mu = 0.99$.

To apply our results to encoded quantum memories, we need to translate a given coherence time $t_c$ or an effective inverse coherence time $\alpha$ into a phase flip probability via $p = (1 - e^{-\alpha})/2$, Eq. (2), convert this into a logical phase flip probability $p_L < 1/2$, and switch back to

$$\alpha_L = -\log(1 - 2p_L).$$

As an example, an initial, physical phase flip probability of 1% is achieved with a coherence time of $t_c = 0.1$ s and an elementary segment length of $L_0 \approx 420$ m or with $t_c = 10$ s and $L_0 \approx 42$ km, whereas a phase flip probability of 5% corresponds to $t_c = 0.1$ s and $L_0 \approx 2193$ m or $t_c = 10$ s and $L_0 \approx 219$ km. The reason for this is that $L_0$ determines both the time each entanglement distribution attempt takes and the success probability of a single attempt. A shorter segment length $L_0$ increases the success probability of each entanglement distribution attempt and decreases the time each entanglement distribution attempt takes. Hence, on average, the quantum memory decays for a shorter period of time and experiences a smaller phase flip probability. Since $p_L$ is supposed to be the logical phase flip probability per time step $\tau$, in an experiment, one needs to error-correct the dephasing quantum memory after every single entanglement distribution attempt in the neighboring segment: when the attempt fails, the stored memories are recovered by an error correction step and continue their storage, whereas, when the attempt succeeds, the two logical memories are ready for a logical Bell measurement. In principle, syndrome detection and Bell measurement could be eventually done in one step after the complete storage [51]. This has the benefit of a huge experimental simplification, but the drawback of an accumulation of dephasing that the error correction code may no longer handle. Moreover, for the statistical rate analysis, with the latter method, the logical phase flip probability would have to be treated jointly with the random variable of the total dephasing, rendering the secret key rate calculation more complicated. Our efficient adaptive syndrome identification switching between dephasing and depolarizing errors would also no longer be applicable. The depolarizing errors due to imperfect initial state preparation and faulty Bell measurement gates are not time-dependent, so, on their own, they are well suited to be detected during the entanglement swapping step. Therefore here, throughout, we employ the frequent stepwise error correction method for the memory dephasing supplemented by a single and separate detection step for the depolarizing errors.

### 1. Two-segment repeater

We consider the two-segment quantum repeater to gain some general insights. In Fig. 11, the secret key rates of both an encoded and an unencoded quantum repeater without memory cutoff are displayed. The five-qubit code outperforms the Steane code, which is unsurprising given that the five-qubit code exhibits lower logical error probabilities both for single-qubit dephasing and two-qubit depolarization. For a coherence time as high as ten seconds, the five-qubit code secret key rate can stay as high as the raw rate up to a distance of $700-800$ km. Such a coherence time may not only be achieved by searching for an appropriate physical system but also by multiplexing techniques [15].

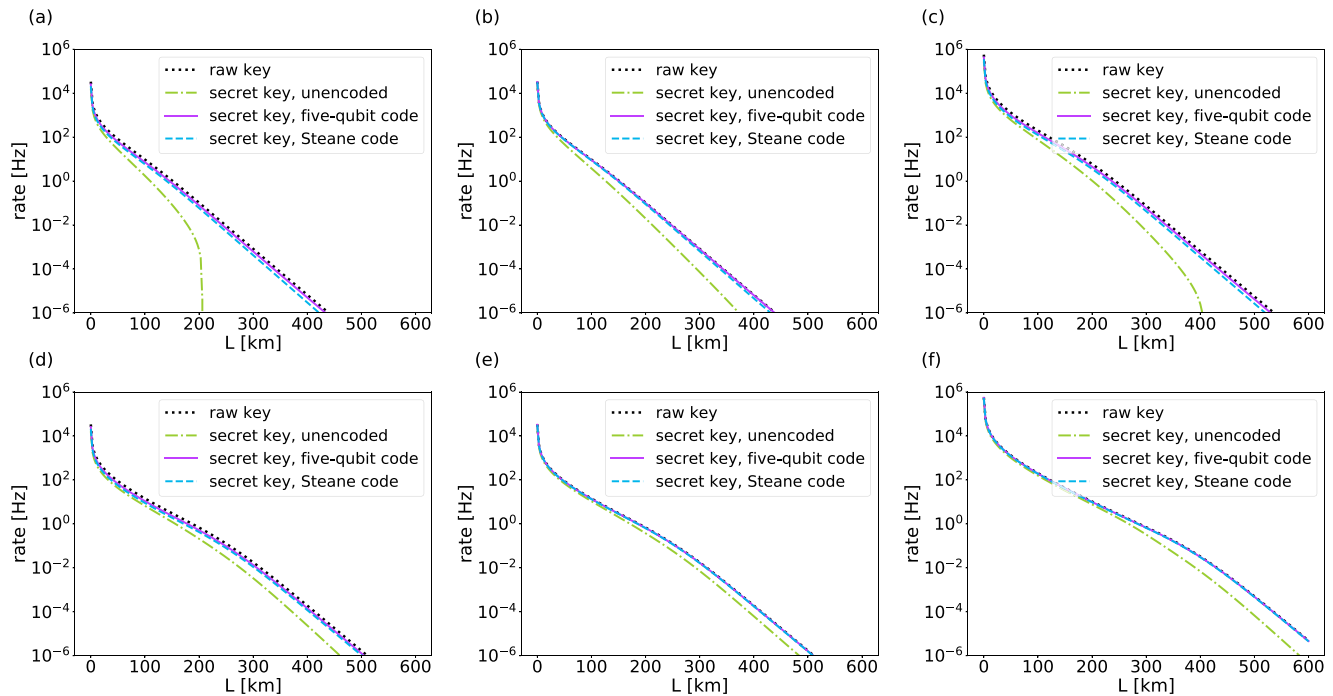

FIG. 12. Raw and secret key rates of a two-segment quantum repeater of length $L$ with memory cutoff after $m$ entanglement distribution attempts for different coherence times $t_c$, zero-distance link-coupling efficiencies $p_0$, Bell state depolarization parameters $\mu$, and initial depolarizations $\mu_0 = \mu$. The subplots correspond to the experimental parameters: (a) $t_c = 0.1$ s, $p_0 = 0.05$, $\mu = 0.97$, $m = 100$; (b) $t_c = 0.1$ s, $p_0 = 0.05$, $\mu = 0.99$, $m = 100$; (c) $t_c = 0.1$ s, $p_0 = 0.7$, $\mu = 0.97$, $m = 50$; (d) $t_c = 10$ s, $p_0 = 0.05$, $\mu = 0.97$, $m = 3000$; (e) $t_c = 10$ s, $p_0 = 0.05$, $\mu = 0.99$, $m = 3000$; and (f) $t_c = 10$ s, $p_0 = 0.7$, $\mu = 0.99$, $m = 5000$.

The raw and secret key rates never exceed 1 MHz since this is the repetition rate $\tau_{\text{clock}}^{-1}$ that we assume. The choice of $p_0$ determines the key rate at zero distance. Additionally, the secret key rate offset depends on the parameter $\mu$, which will become apparent when analyzing the four and eight-segment repeaters. The memory coherence time and the depolarization parameter determine how long the secret key rate can keep up with the raw rate instead of dropping to zero. This is reasonable since fewer errors due to quantum memory storage, measurements, or state preparation ensure an enhanced secret key fraction.

As soon as we start using a so-called memory cutoff [52], interrupting a round of entanglement distribution attempts after a pre-specified number of trials $m$, the benefit of encoding seems to vanish, see Fig. 12, in the sense that the difference between the unencoded and encoded secret key rates decreases. For all considered experimental parameters and distances, the five-qubit and Steane code secret key rates now (almost) coincide with the raw rate, which is, however, reduced compared to an encoded quantum repeater without memory cutoff. Including the memory cutoff improves the secret key fraction at the cost of reducing the raw rate. For memories, encoded with the five-qubit code, the cutoff undermines the benefit of having quantum error correction at hand and reduces the achievable raw rates and, thus, secret key rates. When using the Steane code, the memory cutoff improves the secret key rates for small coherence times of

$t_c = 0.1$ s, whereas, for long coherence times of ten seconds, it rather impairs them. In total, the memory cutoff hinders the secret key distribution for distances and experimental parameters, where the encoded secret key rate can already be kept close to the raw rate thanks to quantum error correction. This behavior is also expected from quantum repeaters with more segments for which we do not have explicit expressions for the probability-generating function of the dephasing variable with a cutoff in the optimal scheme.

In the exact rate analysis of Ref. [15], the minimal $\mu = 1 - p$ values to achieve a nonzero secret key fraction were determined. Setting $\mu_0 = \mu$, that value is 0.920 for a two-segment quantum repeater and BB84 quantum key distribution [53] like here. Since the five-qubit code provides a benefit starting from around an error probability of $p \approx 0.1$, so $\mu \approx 0.9$, we expect to improve the minimal $\mu$ value. Indeed, it is sufficient to have a physical $\mu$ of 0.914 to achieve an effective $\mu$ of 0.920 with the five-qubit code. Since the Steane code provides a benefit only from around $\mu \approx 0.95$, we cannot improve the tolerated $\mu$ with it.

### 2. Four-segment repeater

A quantum repeater with four segments allows to achieve nonzero secret key rates for greater distances, as demonstrated in Fig. 13. Now nonunit $\mu$ and $\mu_0$ values have a stronger impact on the quantum bit error rates, Eq. (A6) in Appendix A, that scale exponentially with the number of segments. This becomes visible in a gap between the raw and secret key rates opening up even at small quantum repeater lengths when considering $\mu = 0.97$ instead of $\mu = 0.99$. The depolarization parameters $\mu$ and $\mu_0$ determine whether the secret key rates follow closely the raw rate before dropping to zero, as it is the case for $\mu = 0.99$ or in parallel with some significant distance as for $\mu = 0.97$. Fortunately, this rather large gap between the unencoded secret key and the raw rate can be overcome with an encoded quantum repeater for $\mu = 0.99$ or at least heavily reduced for $\mu = 0.97$. This shows that the encoded quantum repeater can not only extend the achievable distance for quantum key distribution but also provide higher secret key rates.

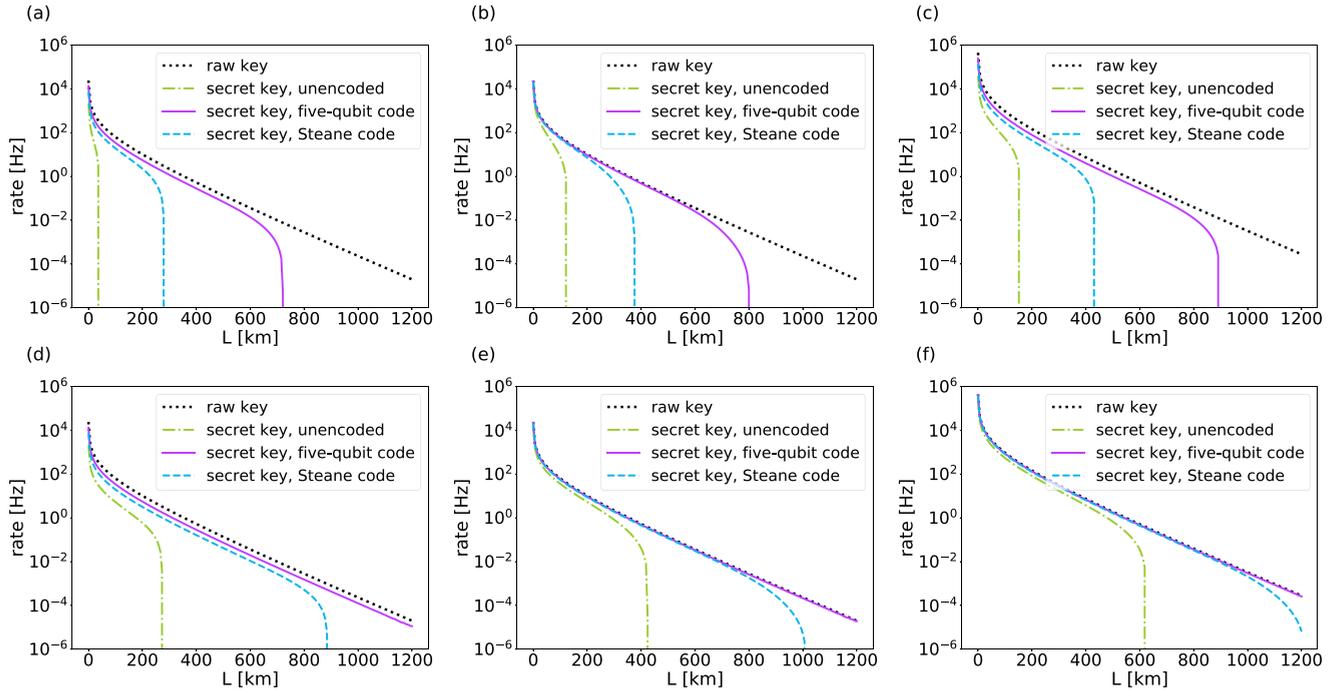

FIG. 13. Raw and secret key rates of a four-segment quantum repeater of length $L$ without memory cutoff for different quantum memory coherence times $t_c$, zero-distance link-coupling efficiencies $p_0$, Bell state depolarization parameters $\mu$, and initial depolarizations $\mu_0 = \mu$. The subplots correspond to the experimental parameters: (a) $t_c = 0.1$ s, $p_0 = 0.05$, $\mu = 0.97$; (b) $t_c = 0.1$ s, $p_0 = 0.05$, $\mu = 0.99$; (c) $t_c = 0.1$ s, $p_0 = 0.7$, $\mu = 0.97$; (d) $t_c = 10$ s, $p_0 = 0.05$, $\mu = 0.97$; (e) $t_c = 10$ s, $p_0 = 0.05$, $\mu = 0.99$; (f) $t_c = 10$ s, $p_0 = 0.7$, $\mu = 0.99$.

For the four-segment repeater with $\mu_0 = \mu$, the minimally required $\mu = 0.965$ for a nonzero secret key rate in an unencoded repeater [15] (for BB84 quantum key distribution like here) is now lowered by around 2% to $\mu \approx 0.945$ if using the five-qubit code. With the Steane code, we can improve the tolerated $\mu$ of the four-segment repeater marginally from 0.965 to 0.959.

### 3. Eight-segment repeater

A quantum repeater with eight segments allows the distribution of secret keys over even further distances, see Fig. 14. Again, we consider the minimal $\mu = 1 - p$ value that is required for a nonzero secret key fraction, setting $\mu_0 = \mu$. This is 0.984 for the unencoded eight-segment repeater with BB84 [15]. Employing the five-qubit code, it suffices to have $\mu = 0.964$, and for the Steane code $\mu \approx 0.973$. Since with $\mu = 0.97$ and $\mu_0 = \mu$ only the five-qubit code would produce nonzero secret key rates, we increased the lower depolarization parameter in Fig. 14 to $\mu = 0.975$.

With $\mu = \mu_0 = 0.975$, the encoded quantum repeater is able to provide nonzero secret key rates in contrast to the unencoded quantum repeater. Nonetheless, the gap between the raw rates and the encoded secret key rates is now more prominent than for the four-segment quantum repeater. As previously, this gap is smaller for the five-qubit code than for the Steane code and can be closed by improving the Bell state measurement and initial depolarization parameter to $\mu = \mu_0 = 0.99$. For a zero-distance link-coupling efficiency of $p_0 = 0.7$, we obtain at 800 km secret key rates of 2.19 Hz in Fig. 14(c) and 4.85 Hz in Fig. 14(f) compared to zero and 1.25 Hz of the unencoded repeater, respectively, so that we even beat ideal twin-field quantum key distribution. For the latter, we assume repetition rates of $\tau^{-1} = \tau_{\text{clock}}^{-1} = 1$ GHz, and we calculated the secret key rate at a distance of 800 km as 0.71 Hz using the secret key fraction formula provided in Ref. [54].

The fact that the secret key rate with encoding manages to keep up with the raw rates better than the unencoded repeater and for greater distances means that increasing the raw rate might be a promising approach to achieving high secret key rates. Since we already distribute entanglement in parallel in the optimal scheme, the raw rates might be further increased by splitting up the total distance into more than eight segments to decrease classical communication times.

### C. Larger quantum repeaters

To determine whether the secret key rate can reach the limits set by the inverse classical communication time units $c_f/L_0$ and light-matter coupling of MHz order, we consider quantum repeaters with more than eight segments for a total length of 800 km. For large elementary segment lengths $L_0$, so small segment numbers $N$, the raw rate is predominantly determined by classical communication times, whereas for small segment lengths, i.e., large segment numbers, the light-matter interface becomes increasingly important so that for $N = 8000$, the raw rate almost approaches the 1 MHz limit set by $\tau_{\text{clock}}$ in Eq. (5).

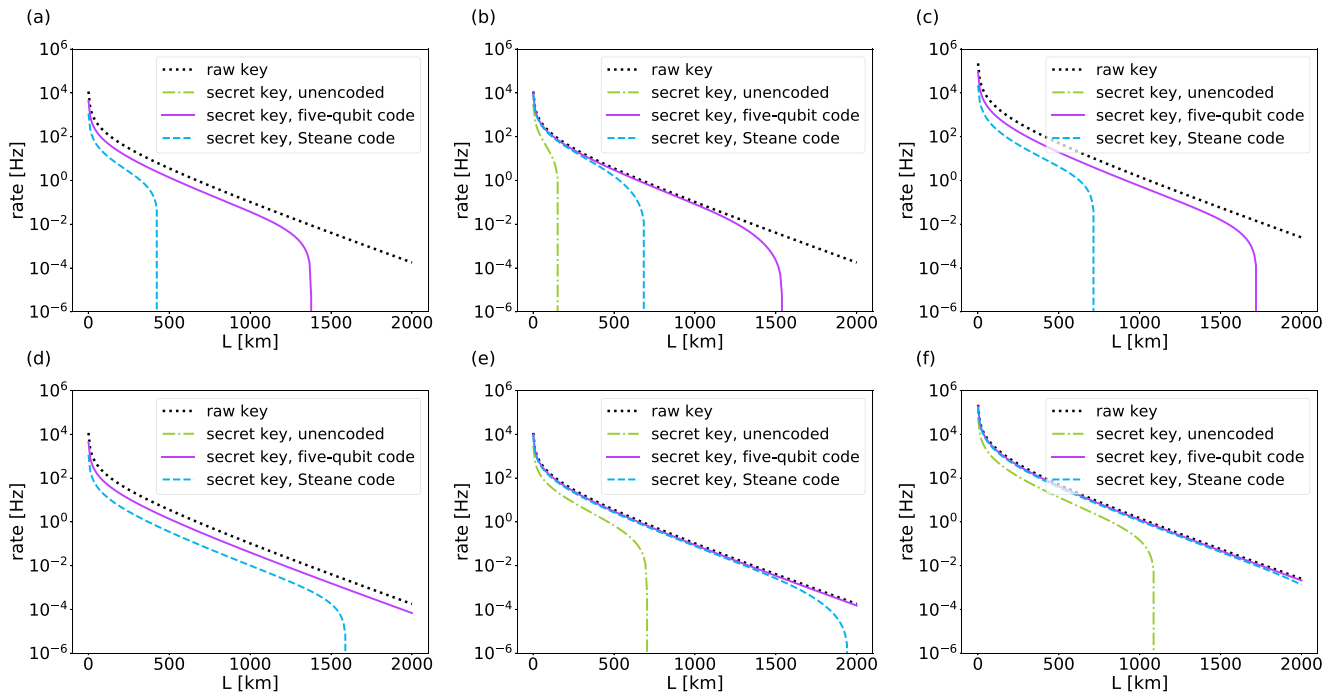

FIG. 14. Raw and secret key rates of an eight-segment quantum repeater without memory cutoff for different quantum memory coherence times $t_c$, zero-distance link-coupling efficiencies $p_0$, Bell state depolarization parameters $\mu$, and initial depolarizations $\mu_0 = \mu$. The subplots correspond to the experimental parameters: (a) $t_c = 0.1$ s, $p_0 = 0.05$, $\mu = 0.975$; (b) $t_c = 0.1$ s, $p_0 = 0.05$, $\mu = 0.99$; (c) $t_c = 0.1$ s, $p_0 = 0.7$, $\mu = 0.975$; (d) $t_c = 10$ s, $p_0 = 0.05$, $\mu = 0.975$; (e) $t_c = 10$ s, $p_0 = 0.05$, $\mu = 0.99$; (f) $t_c = 10$ s, $p_0 = 0.7$, $\mu = 0.99$.

The raw rates can be computed exactly for the optimal scheme or, in general, schemes with entanglement distribution in parallel. However, the dephasing statistics becomes increasingly complex with higher segment numbers for entanglement swapping as soon as possible, unless we make the simplifying and restrictive assumption that Alice and Bob measure their qubits immediately [15,55]. Nonetheless, here, as a very simple lower bound on the secret key fraction of the optimal scheme, we shall consider the fully sequential scheme as proposed in Ref. [15]. This approach is based on the observation that for all $N \leqslant 8$, the dephasing rate was higher in the fully sequential scheme than in the optimal scheme. Extrapolating this behavior to larger repeaters with $N > 8$, the secret key fraction in the fully sequential scheme serves as a lower bound for that in the optimal scheme.

The resulting secret key rates are displayed in Table I. Due to the high segment numbers, the encoded secret key rates show no difference whether we set $t_c = 10$ s or $t_c = 0.1$ s for equal depolarization parameter $\mu$. Hence, we can lower experimentally required memory coherence times and consider $t_c = 0.1$ s and $t_c = 0.001$ s = 1 ms.

For $N = 80$, the unencoded repeater has nonzero secret key rates only for $\mu = 0.999$ and $t_c = 0.1$ s, however, it is one order of magnitude below the raw rate limit of 3.8 MHz, which is approachable both with the five-qubit and the Steane code for the same experimental parameters. The five-qubit code can even achieve $\sim$0.1 MHz for our worst considered experimental parameters of $\mu = 0.99$, $t_c = 1$ ms. In the case of $N = 800$, both the five-qubit and the Steane code approach the raw rate limit, whereas the unencoded repeater cannot produce secret keys for $N = 800$ or 8000.

The higher the segment numbers are, the lower we can allow the memory coherence time to be, which is visible in the five-qubit code secret key rates depending on the two different memory coherence times for $N = 80$ but not anymore for $N = 800$ and 8000. Also, the Steane code produces secret keys for $N = 800$, $\mu = 0.999$, and $t_c = 1$ ms but not for $N = 80$ and otherwise equal parameters. However, the experimental requirements on $\mu$ become increasingly demanding for large repeaters so that, for $N = 8000$, the Steane code fails to produce a secret key even for $\mu = 0.999$ and the five-qubit code secret key rates stay one order of magnitude below the raw rate limit of $\sim$1 MHz. As a result, the secret key rates of the five-qubit code are similar for $N = 800$ and $N = 8000$, making the experimental cost of having 8000 repeater stations too high for gaining almost no benefit. To approach the raw rate limit for $N = 8000$, we need to require physical $\mu = 0.9999$ as well as encoded quantum memories. Then, the five-qubit code produces secret keys of >944 kHz irrespective of $t_c$ and the Steane code achieves >806 kHz for $t_c = 1$ ms and >928 kHz for $t_c = 0.1$ s.

In general, the better our physical error parameters are, the better we can effectively improve them on the logical level after quantum error correction. In Fig. 15, this is shown for the five-qubit and Steane codes, subject to two-qubit depolarization with very low depolarization probabilities $p$. For both codes, we display the higher effective error probability curve—the error parameters corresponding to single logical errors—which is also used to approximate logical two-qubit depolarization in our worst-case treatment. It is visible that a physical two-qubit depolarization probability of $p = 10^{-2} = 0.01$, so $\mu = 0.99$, is improved by about one order of magnitude. A physical $\mu$ of 0.999 is already improved by about two orders of magnitude and $\mu = 0.9999$ by about three.

TABLE I. Secret key rates $S(\mu, t_c)$ for large quantum repeaters consisting of $N = 80$, 800, or 8000 segments at a fixed total length of 800 km for different memory coherence times $t_c$ and depolarization parameters $\mu = \mu_0$. The raw rate is computed for the parallel-distribution scheme, whereas the secret key fraction is calculated with the fully sequential scheme where the dephasing variable statistics is easy to compute. We assume throughout perfect zero-distance link-coupling efficiencies, so $p_0 = 1$, and two dephasing quantum memories per entanglement distribution attempt.

| $N$ | 80 | 800 | 8000 |
|---|---|---|---|
| Raw rate | 3.8 kHz | 72.7 kHz | 967.2 kHz |
| Unencoded $S(\mu = 0.99, t_c = 0.001$ s$)$ | 0 | 0 | 0 |
| Unencoded $S(\mu = 0.99, t_c = 0.1$ s$)$ | 0 | 0 | 0 |
| Unencoded $S(\mu = 0.999, t_c = 0.001$ s$)$ | 0 | 0 | 0 |
| Unencoded $S(\mu = 0.999, t_c = 0.1$ s$)$ | >331 Hz | 0 | 0 |
| Five-qubit $S(\mu = 0.99, t_c = 0.001$ s$)$ | >135 Hz | 0 | 0 |
| Five-qubit $S(\mu = 0.99, t_c = 0.1$ s$)$ | >424 Hz | 0 | 0 |
| Five-qubit $S(\mu = 0.999, t_c = 0.001$ s$)$ | >3.0 kHz | >60.6 kHz | >94.1 kHz |
| Five-qubit $S(\mu = 0.999, t_c = 0.1$ s$)$ | >3.7 kHz | >60.6 kHz | >94.1 kHz |
| Steane $S(\mu = 0.99, t_c = 0.001$ s$)$ | 0 | 0 | 0 |
| Steane $S(\mu = 0.99, t_c = 0.1$ s$)$ | 0 | 0 | 0 |
| Steane $S(\mu = 0.999, t_c = 0.001$ s$)$ | 0 | >13.9 kHz | 0 |
| Steane $S(\mu = 0.999, t_c = 0.1$ s$)$ | >3.6 kHz | >53.1 kHz | 0 |

Overall, adding more intermediate quantum repeater stations for a fixed communication distance pays off with a higher achievable raw rate and lower required physical memory coherence times. However, in addition to having a high resource overhead in the case of many segments, the minimal $\mu$ value for nonzero secret key fraction and thus, secret key rates, becomes increasingly challenging. Therefore, with the experimental prerequisites that we have considered, it does not seem advisable to have more than $N = 800$ quantum repeater stations.

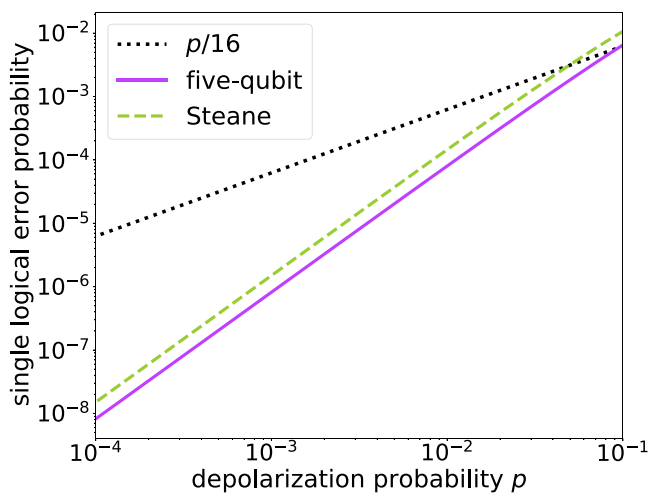

FIG. 15. Single logical error probabilities for the five-qubit and Steane codes, subject to two-qubit depolarization, in comparison to the physical error probability of $p/16$ for each possible two-qubit Pauli error.

## V. CONCLUSION

The main goal of this work was to determine whether logically encoding the quantum memories of a quantum repeater by means of quantum error correction codes can facilitate intrinsically secure quantum key distribution and accomplish higher secret key rates, as compared with memory-assisted quantum communication without the help of quantum error correction. In investigating this, a preceding task that naturally emerged was to translate experimental error probabilities, accounting for imperfect initial states, faulty measurement operations, and finite memory coherence times into new effective logical error probabilities that describe the effective error channels that act on the encoded quantum states. In the secret key rate analysis presented in Ref. [15], the physical and hence experimentally occurring errors are modelled via Pauli noise channels, specifically, single-qubit dephasing to account for quantum memory storage and two-qubit depolarization elsewhere. We have successfully developed a method that imprints the merit of a stabilizer quantum error correction code onto the error parameters of a considered Pauli noise channel, resulting in a logical Pauli noise channel with modified error probabilities. The limiting factors for the computational efficiency of this check matrix model are that all possible Pauli errors and the entire stabilizer group must be computed in the binary representation. Using this formalism, we have demonstrated that memory-corrected quantum repeaters can significantly extend the distances for which quantum key distribution is viable. Additionally, with encoding, secret key rates are enhanced up to the raw rate level for sufficiently high Bell state measurement fidelities and sufficiently well-prepared initial states.

To treat noise on stabilizer states, we have introduced the check matrix model. In contrast to the approach from Ref. [33], serving as an inspiration, we avoided performing calculations in the state picture. Instead, we made use of the stabilizer formalism, so that we only required the stabilizer group generators and logical operators in the binary representation, known from check matrices [5]. This approach proved to be more computationally efficient than the model from Ref. [33], which relies upon error space projectors of exponentially increasing size and number. In the check matrix model, the error space projectors are replaced by binary syndrome vectors. As a result, we were able to consider more complex codes, including those up to eleven qubits. We also generalized our methods to quantum noise channels on two qubits, which are useful to model the Bell state measurements and imperfect initial states in the secret key rate analysis of Ref. [15]. The effective error probabilities obtained in this work are very general and applicable in any given scenario where one is interested in protecting qubits against noise.

Particularly the five-qubit code turned out to be a very promising candidate for quantum memory encoding. This performance largely originated from a specific and for the present application newly introduced notion of adaptive syndrome identification where the measured error syndromes are interpreted in distinct ways depending on the considered quantum noise channel. The Steane code also provided a significant improvement. Overall, the five-qubit and Steane codes led to the best performances in a memory-corrected quantum repeater, while the larger codes that were also considered did not give higher rates despite their higher experimental costs. A subtle but important remark is that we do not deem codes such as the nine-qubit surface code, the Shor code, and the eleven-qubit surface code as not useful per se. We rather cannot treat them in a meaningful manner within the secret key rate analysis [15], which requires noise channels of a certain type. Therefore we encounter a complication in cases where the logical noise channel is no longer of the same type as the physical one. Since we consider averaging or discarding error probabilities as too crude, we sometimes remedy this by assigning all logical error probability values to the worst one so that, in fact, we compute lower bounds on the secret key rate. For the five-qubit and Steane codes, this approximation is not unreasonable since the error probabilities were close to each other from the start. However, for the nine-qubit surface code, the error probabilities greatly vary and simply picking the highest curve would significantly overestimate the low error parameters and, therefore, discard the benefits of the encoding. If we were to average the logical quantum noise channels, the nine-qubit surface code would provide secret key rates somewhere in between the five-qubit and the Steane codes.

A possible approach to more accurately treating also the larger codes would be to generalize the relevant noise in the secret key rate analysis of Ref. [15] to more general Pauli channels. This would allow to consider cases such as for the phase flip or Shor codes, where the logical channel type after, for instance, single-qubit dephasing can become different from the physical one and it may even become a general Pauli noise channel with directed Pauli noise on the logical level. Once an extended model for the quantum key distribution rate analysis is available, the general effective Pauli error parameters for each code could be immediately taken from our effective channel calculations. The corresponding logical error parameters for these general channels for the different codes are presented in Appendix D. Additionally, such an extension would remove the need to approximate a general Pauli noise channel with a two-qubit depolarizing channel. Finally, we note that in the treatment of entanglement distillation and with these first-generation quantum repeaters, it is sometimes useful to consider noisy entangled states with undirected Pauli noise as the input to the distillation procedures. The usual assumption then is that such symmetric states are obtainable from more general noisy states through random local operations. This leads to a very useful simplification at the expense of a suboptimal treatment - very similar to our use of the worst-case logical error parameters in order to obtain sufficiently simple logical Pauli noise channels.

As a result of our analysis, the minimal physical requirements on the experimental parameters for the initial state preparation and Bell state measurement fidelity can be reduced by a few percent when utilizing the five-qubit code with adaptive syndrome identification. Furthermore, nonvanishing secret-key rates up to distances of 2000 km become achievable for high coherence times of ten seconds, which may be accomplished with the help of multiplexing techniques [15]. Using an eight-segment repeater, we even beat ideal twin-field quantum key distribution at a distance of 800 km. For larger quantum repeaters with more intermediate memory stations, we found that the raw rate can only be approached with the help of quantum error correction and especially well when utilizing the five-qubit code. In the case of $N = 80$ or $N = 800$ segments for a fixed communication distance of 800 km and setting $\mu = 0.999$, the five-qubit code was able to practically maintain secret key rates at the raw-rate level even for memory coherence times $t_c$ of only 1 ms (the Steane code achieves this for $t_c = 0.1$ s). However, due to the increasing requirements imposed on the depolarization parameter $\mu$ for a nonzero secret key fraction for growing $N$, the secret key rate did not gain an order of magnitude when moving from $N = 800$ segments to $N = 8000$. To achieve the raw rate limit of $\sim 1$ MHz for $N = 8000$, one would need encoded quantum memories and a physical $\mu$ of 0.9999.

The logical error parameters should encompass all experimental circumstances, including the potential requirement of additional faulty operations for quantum error correction, such as state preparation or syndrome detection. Quantifying these additional effects in future research would be intriguing. It would also be worthwhile investigating the possibility of reconstructing the error syndrome from the outcomes of the transversal Bell state measurements, thereby avoiding additional error sources. Another remaining task is to establish a concrete protocol for encoded entanglement distribution to provide practical guidelines for experimental realization and study its impact on the distribution probability. Our findings demonstrate that encoding can lower the minimum required fidelity of Bell state measurements for achieving nonzero secret key rates, provided that the effects of additional faulty operations and potentially lower initial distribution probability are negligible. Indeed our use of relatively small instances of quantum error correction codes appears to be a good choice with regards to such trade-offs, combining a relatively modest resource overhead with reasonable initial state preparation probabilities and a suppression of propagating errors that otherwise are more likely to occur in larger encoding circuits.

Our research focused on stabilizer codes, also known as additive codes, which are just one of many options for encoding quantum memories. Exploring quantum codes beyond the scope of stabilizer codes, such as low-density parity-check codes [56] or nonadditive codes [57], for memory-corrected quantum repeaters, is certainly of further interest. The logical error rates of quantum low-density parity check codes in dependence on physical error rates have been recently investigated with Monte Carlo simulations and an experimental realization of these codes on reconfigurable atom arrays has been proposed [58].

The developed check matrix model can deal with Pauli noise channels, which was sufficient in the context of the exact rate analysis of Ref. [15] that modelled errors via depolarization and dephasing. Nonetheless, generalizing our formalism to general quantum noise channels would be interesting to have means of treating, for instance, the amplitude damping channel [5]. However, provided Pauli noise affects a physical system, our results are applicable not only for quantum communication but also in the context of quantum computation.

## ACKNOWLEDGMENTS

We thank Frank Schmidt for useful input. We further acknowledge funding from the BMBF in Germany (QR.X, QuKuK) and from the Deutsche Forschungsgemeinschaft (DFG, German Research Foundation)–Project-ID 429529648–TRR 306 QuCoLiMa ("Quantum Cooperativity of Light and Matter").

## APPENDIX A: SECRET KEY RATE ANALYSIS

We briefly review the secret key rate analysis along the lines of the treatment and model given in Ref. [15]. The model relies on two central quantities: first, the total waiting time needed to distribute an entangled state between the two communicating parties, and second, the dephasing time that characterizes the quality of the distributed state and is determined by the decoherence of the quantum memories occurring while they wait for a neighboring segment to become ready for entanglement swapping. These two random variables were computed with the help of probability-generating functions.

### 1. The error model and final shared noisy state

For the simple case of a two-segment quantum repeater, the states $\rho_{12}$ and $\rho_{34}$ are distributed in the two respective segments of length $L_0$, see Fig. 16. If $p$ is the probability that a distribution attempt succeeds, the number of required attempts is a geometrically distributed random variable $\mathcal{N}_1$ for the first segment and $\mathcal{N}_2$ for the second segment. To achieve entanglement swapping $\mathcal{S}$, a Bell state measurement on the qubits 2 and 3 is performed resulting upon success in an entangled state of qubits 1 and 4,

$$\rho_{14} = \mathcal{S}(\rho_{1234}) = \frac{{}_{23}\langle\Psi^+|\mathcal{B}_{\mu,23}(\rho_{1234})|\Psi^+\rangle_{23}}{\mathrm{Tr}({}_{23}\langle\Psi^+|\mathcal{B}_{\mu,23}(\rho_{1234})|\Psi^+\rangle_{23})}. \quad (A1)$$

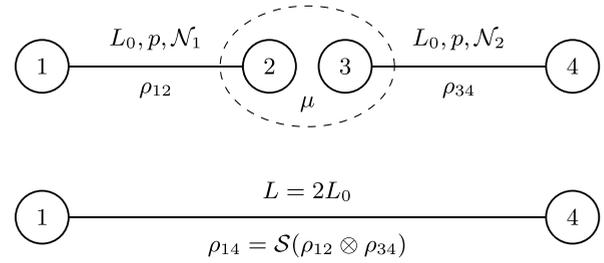

FIG. 16. Two-segment quantum repeater that distributes the state $\rho_{14}$ to the end users, separated by a distance of $L = 2L_0$, via entanglement swapping $\mathcal{S}$ on the states $\rho_{12}$ and $\rho_{34}$. An attempt to distribute the states $\rho_{12}$ and $\rho_{34}$ over their respective segments of length $L_0$ succeeds with probability $p$. The total numbers of attempts until success $\mathcal{N}_1$ and $\mathcal{N}_2$ are geometrically distributed so the probability of exactly $i$ attempts is given by $p(1-p)^{i-1}$. The figure is recreated with slight adaptations from Ref. [15].

To take into account imperfect state preparation and distribution, the fidelity parameter $F_0$ can describe some initial dephasing and $\mu_0$ an initial depolarization so that the initial state is given by

$$\rho_0 = \mathcal{B}_{\mu_0}(F_0|\Psi^+\rangle\langle\Psi^+| + (1-F_0)|\Psi^-\rangle\langle\Psi^-|). \quad (A2)$$

For the secret key rate analysis, we will set $F_0 = 1$ and only consider initial depolarization to model imperfect states. The action of $\mathcal{M}_\alpha$, Eq. (3), is equal on the first and the second qubit of $\rho_0$ so either will be denoted by $\mathcal{M}_\alpha$. It can now be shown that

$$\mathcal{M}_\alpha(\rho_0) = \mathcal{B}_{\mu_0}(F|\Psi^+\rangle\langle\Psi^+| + (1-F)|\Psi^-\rangle\langle\Psi^-|)$$

with the modified fidelity given by

$$F = \tfrac{1}{2}(2F_0 - 1)e^{-\alpha} + \tfrac{1}{2}.$$

For two states $\rho_{12}$ and $\rho_{34}$ of the form of Eq. (A2) with their respective initial fidelities $F_1$ and $F_2$ as well as initial depolarization parameters $\mu_1$ and $\mu_2$, evaluating Eq. (A1) results in a swapped state

$$\rho_{14} = \mathcal{B}_{\mu_d}(F_d|\Psi^+\rangle\langle\Psi^+| + (1-F_d)|\Psi^-\rangle\langle\Psi^-|),$$

where $\mu_d = \mu\mu_1\mu_2$ and

$$F_d = \tfrac{1}{2}(2F_1 - 1)(2F_2 - 1) + \tfrac{1}{2}.$$

Thus the form of the state is preserved, which is also the case for larger repeaters. Generalizing to an $N$-segment quantum repeater, the final distributed state becomes

$$\rho_N = \mathcal{B}_{\mu_N}\left(\frac{1 + (2F_0 - 1)^N e^{-\alpha D_N}}{2}|\Psi^+\rangle\langle\Psi^+| + \frac{1 - (2F_0 - 1)^N e^{-\alpha D_N}}{2}|\Psi^-\rangle\langle\Psi^-|\right), \quad (A3)$$

whereby $\mu_N = \mu_0^N \mu^{N-1}$ and the random variable $D_N(\mathcal{N}_1, \dots, \mathcal{N}_N)$ describes the total number of time units contributing to the total dephasing, e.g., $D_2(\mathcal{N}_1, \mathcal{N}_2) = |\mathcal{N}_1 - \mathcal{N}_2|$. The dephasing variable $D_N$ determines the final state quality and is essential for the extractable secret key fraction.

The total waiting time random variable $W_N$, in turn, determines the achievable raw communication rate. Note that these variables strongly depend on the distribution and swapping schemes. For instance, with parallel distribution of entangled photons in each segment the waiting time $W_N = \max(\mathcal{N}_1, \ldots, \mathcal{N}_N)$ would be minimized compared to a waiting time of $W_N = \mathcal{N}_1 + \cdots + \mathcal{N}_N$ for sequential distribution. However, using parallel distribution schemes potentially accumulates dephasing time due to the increased parallel storage in quantum memories compared to $D_N = \mathcal{N}_2 + \cdots + \mathcal{N}_N$ for sequential distribution. Both variables $D_N$ and $W_N$ together determine the secret key rate of the quantum repeater. Their statistics such as expectation values are computed using probability-generating functions.

### 2. The secret key rate

From the noisy state shared between the two end-node users, one can calculate the achievable asymptotic secret key fraction in the BB84 protocol [53] for one-way postprocessing [59]

$$r = 1 - h(\bar{e}_x) - h(\bar{e}_z) \tag{A4}$$

with $\bar{e}_x$ and $\bar{e}_z$ being the averages of the quantum bit error rates,

$$e_z = \langle 00|\rho_n|00\rangle + \langle 11|\rho_n|11\rangle,$$
$$e_x = \langle +-|\rho_n|+-\rangle + \langle -+|\rho_n|-+\rangle, \tag{A5}$$

where $\rho_N$ is the final distributed state in an $N$-segment quantum repeater, and the binary entropy is given by

$$h(p) = -p\log_2(p) - (1-p)\log_2(1-p).$$

The quantum bit error rates, Eq. (A5), of the final distributed state $\rho_N$ in Eq. (A3) are as follows:

$$e_z = \tfrac{1}{2}\left(1 - \mu^{N-1}\mu_0^N\right),$$
$$e_x = \tfrac{1}{2}\left(1 - \mu^{N-1}\mu_0^N(2F_0-1)^N e^{-\alpha D_N}\right). \tag{A6}$$

This means that to evaluate the average quantum bit error rates, one needs to compute the expectation value of $e^{-\alpha D_N}$. Afterwards, $\bar{e}_z$ and $\bar{e}_x$ can be inserted into the formula for the secret key fraction, Eq. (A4).

The ingredient currently missing to convert this into a secret key rate is the raw rate $R$, determined by the inverse of the average quantum repeater waiting time $T$. If $W$ is the number of time steps required to distribute one entangled photonic qubit pair over the entire communication distance, $T$ is simply given as the expectation value of $W$, multiplied by a unit that characterizes the time necessary for each attempt. Since we need classical communication to confirm successful distribution before we can initiate entanglement swapping, this is just the elementary time unit of a quantum repeater, so $\tau = \tau_{\text{clock}} + L_0/c_f$ which was introduced in Eq. (5). The secret key rate is then given by

$$S = Rr = \frac{r}{T}.$$

To maximize it, one needs a scheme with an optimal balance between small dephasing $D_N$, so large $e^{-\alpha D_N}$, and a small waiting time $W_N$. Reference [15] focuses on fast parallel distribution schemes to minimize the total waiting time. To reduce the dephasing time among such parallel distribution schemes the authors perform entanglement swapping as soon as two neighboring segments become ready (and show for $N = 2, \ldots, 8$ and conjecture for $N > 8$ that this indeed leads to a globally optimized dephasing time). We will refer to this scheme as the optimal scheme since its optimality with respect to the secret key rate was shown for the three-segment repeater in all relevant regimes, especially in the limit of improving hardware parameters, and is conjectured for $N > 3$ [15].

### 3. Probability-generating functions

Probability-generating functions [15] provide the means to capture the statistics of a non-negative integer-valued random variable $X$ by working with polynomial functions. The probability-generating function of a random variable $X$ is defined as

$$G_X(t) = \sum_{k=0}^{\infty} P(X=k)t^k = \mathbb{E}[t^X],$$

where $P(X=k)$ is the probability that the random variable $X$ assumes the integer value $k$ and $\mathbb{E}[t^X]$ denotes the expectation value of $t^X$. We will denote the probability-generating function of the dephasing time random variable via $G_D$ and that of the waiting time random variable via $G_W$.

If the probability-generating function of a random variable is explicitly known, we can obtain all statistical information, for instance, the average value $\mathbb{E}[X] \equiv \bar{X} = G_X'(1)$. For $\alpha \geqslant 0$, we have $\mathbb{E}[e^{-\alpha X}] = G_X(e^{-\alpha})$.

The probability-generating function of the waiting time for parallel distribution schemes, $W_N = \max(\mathcal{N}_1, \ldots, \mathcal{N}_N)$, was obtained in Ref. [15], and from that the average waiting time

$$\overline{W}_N = \frac{d}{dt}G_W(t)\Big|_{t=1} = \sum_{i=1}^{N}(-1)^{i+1}\binom{N}{i}\frac{1}{1-q^i}, \tag{A7}$$

where $q = 1 - p$ is the failure probability of one entanglement distribution attempt.

For a two-segment repeater, also memory cutoff to prevent large dephasing has been considered, where the procedure is interrupted after the total storage time has exceeded a pre-specified value. As mentioned before, for the three-segment repeater all possible schemes were analyzed, and it was found that in terms of the secret key rate indeed the parallel-distribution optimal-dephasing scheme outperforms the other schemes in all relevant parameter regimes and especially for improved hardware parameters. The authors conjecture that this remains true also for $N > 3$. The four-segment repeater was studied for the doubling scheme, the iterative scheme, and the optimal swapping scheme [15]. For the eight-segment repeater, the doubling and the optimal scheme as well as three somewhat less important ones were analyzed. The optimal scheme minimizes the dephasing errors compared to the other four schemes which provides further evidence for its optimality, without, however, being a strict proof. Note that the optimal scheme was designed to minimize $\mathbb{E}[D]$, which does not imply that $\mathbb{E}[e^{-\alpha D}]$ is maximized.

To understand the constant $\alpha$ and give the variable $D$ an experimental meaning, note that the number of contributing dephasing time steps $D$ needs to be multiplied with the elementary time unit of the quantum repeater $\tau$ to give the actual passage of time during which dephasing occurs. Now depending on the coherence time $t_c$ of the system, the memory dephasing time $\tau D$ will have a different effect in terms of the errors that have occurred, for instance, if $\tau D \ll t_c$ likely no errors happen. Thus we see that $\alpha$ is an effective inverse coherence time, as it is described by Eq. (4).

For quantum repeaters with $N > 8$ and parallel distribution, the waiting times can be computed analytically with Eq. (A7), whereas there were no exact expressions obtained for the probability-generating function of the dephasing random variable of the optimal scheme due to the increasing complexity in its calculation.

Reference [15] also discusses spatial, temporal, or spectral multiplexing, so using replicas of all memories and channels (potentially interacting with each other), operating a single fiber at high clock rates, or operating it at different wavelengths. Effectively, this enhances the memory coherence time. However, here we are more interested in improving the quantum bit error rates, Eq. (A6), that exhibit an exponential scaling of $\mu$ and $\mu_0$ with the number of segments $N$ prohibiting scaling up the quantum repeater to more segments and larger distances as long as no quantum error correction or detection is included. This is where the check matrix model comes into play and establishes effective error parameters for encoded quantum memories when error correction is performed. This will of course not only improve the depolarization parameters $\mu$ and $\mu_0$, which we are mainly aiming at due to the strong limitations these impose on the achievable secret key rates, but also the effective inverse memory coherence time $\alpha$. Conceptually, we move from a first-generation quantum repeater to a second-generation quantum repeater.

## APPENDIX B: EFFECTIVE QUANTUM NOISE CHANNELS

The model in Ref. [33] provides a framework for obtaining effective quantum noise channels that can be applied to different stabilizer encodings. We shall introduce it with a few more detailed explanations. In the process, it becomes apparent that multiple physical Pauli noise channels, acting on stabilizer code qubits, translate to a logical Pauli noise channel.

The physical qubits are defined on the Hilbert space $\mathcal{H}_1 = \mathbb{C}^2$ with the orthonormal basis $\{|0\rangle, |1\rangle\}$. The Hilbert space of $n$ physical qubits is given by $\mathcal{H}_n = \mathbb{C}^{2\otimes n}$ so to define the logical qubit we need to specify a two-dimensional subspace of $\mathcal{H}_n$, denoted by $V_0$. Its orthonormal basis is written as $\{|0_L\rangle, |1_L\rangle\}$. Hence, one chooses a map

$$|0\rangle \in \mathcal{H}_1 \rightarrow |0_L\rangle \in \mathcal{H}_n, \quad |1\rangle \in \mathcal{H}_1 \rightarrow |1_L\rangle \in \mathcal{H}_n. \quad \text{(B1)}$$

The remaining Hilbert space is then partitioned into orthogonal, two-dimensional subspaces $V_i$, $i \in \{1, \ldots, 2^{n-1} - 1\}$. Ideally, this is done in a way that typical errors, e.g., Pauli errors on one or multiple qubits, map the space $V_0$ to distinct $V_i$.

As an example, the Hilbert space of five qubits is 32-dimensional which implies a partitioning into 16 two-dimensional subspaces. One of these is the logical code subspace and the remaining fifteen subspaces are obtained from the three possible Pauli errors per qubit. Note that since the five-qubit code can correct all single-qubit errors, different single-qubit errors map $V_0$ to distinct error spaces as it is illustrated in Fig. 2.

To calculate the state $|0_L\rangle\langle 0_L|$, we use that a projector on the logical zero is the same as the projector on the $+1$ eigenspace of all $n - 1$ generators of the stabilizer group and the logical $Z$-operator $Z_L$, thus

$$|0_L\rangle\langle 0_L| = \frac{I + Z_L}{2} \prod_{i=1}^{n-1} \frac{I + g_i}{2}. \quad \text{(B2)}$$

This can also be used to compute $|0_L\rangle\langle 1_L| = |0_L\rangle\langle 0_L|X_L$ or $|1_L\rangle\langle 1_L| = X_L|0_L\rangle\langle 0_L|X_L$. Note that we can equivalently represent $|1_L\rangle\langle 1_L|$ by replacing the plus sign in front of $Z_L$ in Eq. (B2) with a minus sign, since $-Z_L|1_L\rangle = |1_L\rangle$.

The error space construction is done with trial and error, starting with all possible single-qubit Pauli errors and then moving to two-qubit Pauli errors and so on until $2^{n-1} - 1$ error spaces are found. This ensures that if an error space is degenerate, the error with the smallest weight is associated with it.

Concretely, calling $E$ a Pauli error under consideration to construct the error space $V_i$, spanned by the orthonormal basis $\{E|0_L\rangle, E|1_L\rangle\}$, we compute the noisy projector

$$P_i = E|0_L\rangle\langle 0_L|E + E|1_L\rangle\langle 1_L|E \quad \text{(B3)}$$

and then check that $P_i P_j = 0 \ \forall j < i$. If now $|a\rangle \in V_i$ with projector $P_i$ and $|b\rangle \in V_j$ with projector $P_j$, we have for their inner product

$$\langle a|b\rangle = \langle a|P_i^\dagger P_j|b\rangle = \langle a|P_i P_j|b\rangle = 0.$$

Therefore elements belonging to different error spaces are indeed orthogonal with this construction. The procedure yields a list of error spaces $V_i$ and associated errors $E_i$. To correct errors on the logical qubit, we will assume that it is measured and, thereby, projected onto one of the eigenspaces $V_i$. Since Pauli errors are self-inverse, the recovery is then simply achieved by applying the associated error $E_i$ to the noisy logical qubit.

Usually, one deals with more errors than available subspaces and we cannot correct every error but only a subset of all possible errors. Errors outside of that subset induce logical errors. Still, a good code should reduce the effective error on the logical level compared to the possible errors on one physical qubit. If a single-qubit channel $\mathcal{E} : \mathcal{D}(\mathcal{H}_1) \rightarrow \mathcal{D}(\mathcal{H}_1)$, where $\mathcal{D}(\mathcal{H})$ is the set of all density operators on $\mathcal{H}$, acts on all $n$ physical qubits that compose one logical qubit, we aim at replacing the noise map $\mathcal{E}^{\otimes n}$ by an effective map $\mathcal{E}_L : \mathcal{D}(V_0) \rightarrow \mathcal{D}(V_0)$. In this way, the merit of the encoding is imprinted in the modified map $\mathcal{E}_L$, and one can use already known results for $|\psi\rangle \in \mathcal{H}_1$ for the encoded equivalent $|\psi_L\rangle \in \mathcal{H}_1^{\otimes n}$.

There will be two restrictions. First, we consider only Pauli noise channels which for a single-qubit $\rho \in \mathcal{D}(\mathcal{H}_1)$ are written as

$$\mathcal{E}(\rho) = \sum_{j=0}^{3} \lambda_j \sigma_j \rho \sigma_j, \quad \text{(B4)}$$

where $\sigma_0 = I$, $\sigma_1 = X$, $\sigma_2 = Y$, $\sigma_3 = Z$, and $\sum_{j=0}^{3} \lambda_j = 1$. Note that this includes the depolarizing channel and the dephasing channel. Second, we only consider stabilizer codes, so the logical zero and one, $|0_L\rangle$ and $|1_L\rangle$, are always stabilizer states.

After applying the overall noise channel $\mathcal{E}^{\otimes n}$ to an arbitrary state $\rho \in \mathcal{D}(V_0)$, projecting onto a $V_i$, renormalizing, and then rotating back to $V_0$, one has a quantum state in $\mathcal{D}(V_0)$ which can be written as

$$\mathcal{E}_L^i(\rho) = \sum_{j,k=0}^{3} \lambda_{j,k}^i \sigma_j^L \rho \sigma_k^L, \tag{B5}$$

whereby $\sigma_j^L : V_0 \to V_0$ are logical Pauli operators. By construction, this single-qubit Pauli channel $\mathcal{E}_L^i$ is trace-preserving and completely positive. The $\lambda_{j,k}^i$ are effective noise parameters and depend on the encoding from Eq. (B1), on the original error parameters $\lambda_j$ of the quantum noise channel in Eq. (B4), and on the error space $V_i$ that one projects onto.

Now we use the Choi-Jamiolkowski isomorphism [60] between quantum channels, described by completely positive maps, and higher-dimensional quantum states, described by density matrices. For this, we define the maximally entangled state

$$|\Phi^+\rangle = \frac{1}{\sqrt{2}} (|0_L\rangle \otimes |0\rangle + |1_L\rangle \otimes |1\rangle) \in V_0 \otimes \mathcal{H}_1,$$

first apply $\mathcal{E}^{\otimes n} \otimes I_{\mathcal{H}_1}$ onto $|\Phi^+\rangle\langle\Phi^+|$, then $P_i \otimes I_{\mathcal{H}_1}$ on the resulting state to project it onto an error space $V_i$, and subsequently rotate back to $V_0$ with $E_i$. This procedure yields a state that is isomorphic to the channel $\mathcal{E}_L^i$ from Eq. (B5).

For $|0_L\rangle$ and $|1_L\rangle$ being stabilizer states, we get the following pattern from this procedure [33]:

$$|0_L\rangle\langle 0_L| \otimes |0\rangle\langle 0| \to (a_0^i|0_L\rangle\langle 0_L| + b_0^i|1_L\rangle\langle 1_L|) \otimes |0\rangle\langle 0|,$$
$$|1_L\rangle\langle 1_L| \otimes |1\rangle\langle 1| \to (a_1^i|1_L\rangle\langle 1_L| + b_1^i|0_L\rangle\langle 0_L|) \otimes |1\rangle\langle 1|,$$
$$|0_L\rangle\langle 1_L| \otimes |0\rangle\langle 1| \to (c_0^i|0_L\rangle\langle 1_L| + d_0^i|1_L\rangle\langle 0_L|) \otimes |0\rangle\langle 1|,$$
$$|1_L\rangle\langle 0_L| \otimes |1\rangle\langle 0| \to (c_1^i|1_L\rangle\langle 0_L| + d_1^i|0_L\rangle\langle 1_L|) \otimes |1\rangle\langle 0|. \tag{B6}$$

To explain this in a little more detail than in Ref. [33], consider $|0_L\rangle\langle 0_L|$ from Eq. (B2). Due to error correction, we either have terms in the total operation that are (up to prefactors representing their occurrence probabilities) in the stabilizer group—in these cases, recovery was successful—or that commute with the stabilizer group but are not in it, so where we have introduced a logical error. Since both cases commute with the generators $g_1, \ldots, g_{n-1}$, it essentially comes down to what happens at $Z_L$. Now $P_i$ from Eq. (B3) by inspection trivially commutes with $Z_L$ (and with $X_L$ which becomes relevant when considering $|0_L\rangle\langle 1_L|$). Apart from that we only deal with Pauli matrices in the total operation which can only commute or anticommute with $Z_L$. Thus the sign in front of $Z_L$ either remains as it is, maintaining the state $|0_L\rangle\langle 0_L|$, or becomes $-Z_L$, changing the overall state to $|1_L\rangle\langle 1_L|$.

For symmetry reasons—the commutation relations of $Z_L$ or $X_L$ and the terms that constitute the total operation remain equal irrespective of whether we consider $|0_L\rangle\langle 0_L|$ or $|1_L\rangle\langle 1_L|$ and $|0_L\rangle\langle 1_L|$ or $|1_L\rangle\langle 0_L|$—we furthermore have that $a_0^i = a_1^i \equiv a^i$, $b_0^i = b_1^i \equiv b^i$, $c_0^i = c_1^i \equiv c^i$, and $d_0^i = d_1^i \equiv d^i$.

In practice, to determine the transformations of the kind of Eq. (B6), we first compute the necessary density matrix in terms of projectors like in Eq. (B2). Next, we use that a $2^n \times 2^n$ density matrix can be written as

$$\hat{\rho} = \sum_{i,j=1}^{2^n} \rho_{ij} |\mathbb{B}(i-1)\rangle\langle\mathbb{B}(j-1)|,$$

where $\rho_{ij}$ is the matrix entry at position $(i, j)$ and column $j$, and $\mathbb{B}$ is the binary representation of an integer $k$, filling up with zeros if necessary to reach a bit string length of $n$. For appropriate $x_k, y_k \in \{0, 1\}$, we have

$$|\mathbb{B}(i)\rangle\langle\mathbb{B}(j)| = |x_1 \ldots x_n\rangle\langle y_1 \ldots y_n|$$
$$= |x_1\rangle\langle y_1| \otimes \cdots \otimes |x_n\rangle\langle y_n|$$

onto which the noise $\mathcal{E}^{\otimes n}$, the measurement $P_i$, and the correction $E_i$ are applied.

In the basis spanned by the set $\{|i_L\rangle \otimes |j\rangle\}_{i,j=0}^{1}$, one obtains from the pattern of Eq. (B6) the density matrix

$$\rho_M = \frac{1}{2} \begin{bmatrix} a^i & 0 & 0 & c^i \\ 0 & b^i & d^i & 0 \\ 0 & d^i & b^i & 0 \\ c^i & 0 & 0 & a^i \end{bmatrix}.$$

Changing to the Bell state basis and renormalizing, we see that $\rho_M$ is diagonal (note that $(X_L \otimes I)|\Phi^+\rangle = |\Psi^+\rangle$, $(Z_L \otimes I)|\Phi^+\rangle = |\Phi^-\rangle$, and $(Y_L \otimes I)|\Phi^+\rangle = i|\Psi^-\rangle$), so that we can read off the effective noise parameters

$$\lambda_0^i \equiv \lambda_{0,0}^i = \frac{1}{2} \frac{a^i + c^i}{a^i + b^i}, \quad \lambda_2^i \equiv \lambda_{2,2}^i = \frac{1}{2} \frac{b^i - d^i}{a^i + b^i},$$
$$\lambda_1^i \equiv \lambda_{1,1}^i = \frac{1}{2} \frac{b^i + d^i}{a^i + b^i}, \quad \lambda_3^i \equiv \lambda_{3,3}^i = \frac{1}{2} \frac{a^i - c^i}{a^i + b^i}, \tag{B7}$$

where $\lambda_{j,k}^i = 0$ for all $j \neq k$ and $p_i = a^i + b^i$ is the probability of landing in the error space $V_i$ after the syndrome measurement.

To take into account all possible outcomes of the error space measurement, we average over all possible error syndrome results. We then have for any $\rho \in \mathcal{D}(V_0)$

$$\mathcal{E}_L(\rho) = \sum_{j,k=0}^{3} \Lambda_{j,k} \sigma_j^L \rho \sigma_k^L,$$

where the mean noise parameters $\Lambda_{j,k}$ are given as the sum of the effective noise parameters $\lambda_{j,k}^i$, weighted by the probability of landing in the error space $V_i$ after applying the noise $\mathcal{E}^{\otimes n}(\rho)$, so

$$\Lambda_{j,k} = \sum_{i=0}^{2^{n-1}-1} p_i \lambda_{j,k}^i. \tag{B8}$$

For stabilizer codes and Pauli noise channels on the physical level, we have $\Lambda_{j,k} = 0$ for all $j \neq k$ and obtain a Pauli noise channel also on the logical level with modified error parameters $\Lambda_i \equiv \Lambda_{i,i}$.

To analyze repetition codes this model is rather convenient since working in the state picture is easily feasible. In the case of $n$ repetitions, we have $|0_L\rangle = |0\rangle^{\otimes n}$ and $|1_L\rangle = |1\rangle^{\otimes n}$ spanning the code space which can protect against up to $\lfloor n/2 \rfloor$ (so $n/2$ rounded down) bit flip errors. By means of light abuse of our previous notation, we refer to the error spaces $V_i$ as the ones where $i$ errors have occurred which means that we actually have $\binom{n}{i}$ different $V_i$. Applying the Pauli noise channel from Eq. (B4) on the physical qubit $|0\rangle$ results in

$$\mathcal{E}(|0\rangle\langle 0|) = \lambda_0 |0\rangle\langle 0| + \lambda_1 |1\rangle\langle 1| + \lambda_2 |1\rangle\langle 1| + \lambda_3 |0\rangle\langle 0|$$
$$= (\lambda_0 + \lambda_3)|0\rangle\langle 0| + (\lambda_1 + \lambda_2)|1\rangle\langle 1|,$$

so that we get

$$\mathcal{E}^{\otimes n}(|0_L\rangle\langle 0_L|) = [(\lambda_0 + \lambda_3)|0\rangle\langle 0| + (\lambda_1 + \lambda_2)|1\rangle\langle 1|]^{\otimes n}.$$

This shows that every factor of $(\lambda_0 + \lambda_3)$ counts the number of times no error on the physical level has occurred and each factor of $(\lambda_1 + \lambda_2)$ counts one bit flip error. Therefore it follows for $a^i$ and analogously for $b^i$, $c^i$, and $d^i$:

$$a^i = (\lambda_0 + \lambda_3)^{n-i}(\lambda_1 + \lambda_2)^i,$$
$$b^i = (\lambda_0 + \lambda_3)^i(\lambda_1 + \lambda_2)^{n-i},$$
$$c^i = (\lambda_0 - \lambda_3)^{n-i}(\lambda_1 - \lambda_2)^i,$$
$$d^i = (\lambda_0 - \lambda_3)^i(\lambda_1 - \lambda_2)^{n-i}.$$

Taking into account that there are $\binom{n}{i}$ ways to distribute $i$ errors among $n$ qubits, we have $p^i = \binom{n}{i}(a^i + b^i)$ and the logical error rates, so the mean noise parameters, become using Eqs. (B7) and (B8):

$$\Lambda_0 = \frac{1}{2} \sum_{i=0}^{\lfloor n/2 \rfloor} \binom{n}{i}(a^i + c^i) \equiv \lambda_I,$$
$$\Lambda_1 = \frac{1}{2} \sum_{i=0}^{\lfloor n/2 \rfloor} \binom{n}{i}(b^i + d^i) \equiv \lambda_X,$$
$$\Lambda_2 = \frac{1}{2} \sum_{i=0}^{\lfloor n/2 \rfloor} \binom{n}{i}(b^i - d^i) \equiv \lambda_Y,$$
$$\Lambda_3 = \frac{1}{2} \sum_{i=0}^{\lfloor n/2 \rfloor} \binom{n}{i}(a^i - c^i) \equiv \lambda_Z.$$

While this calculation was rather straightforward in case of repetition codes, it is not as simple for more complex codes that involve density matrices and error space projectors of exponentially increasing size. Therefore, in Ref. [33], only the five-qubit code was additionally considered apart from repetition codes. Based on and inspired by this approach, we develop the model presented in Sec. III B that avoids working in the state picture and is, thus, able to handle stabilizer codes more efficiently.

## APPENDIX C: CHECK MATRIX MODEL IMPLEMENTATION

For each error, we need to find its occurrence probability which is determined by the Pauli noise channel under consideration. For a single-qubit depolarizing channel, every possible multiqubit Pauli error can occur, whereby each identity term in the tensor product contributes with $1 - \frac{3}{4}p$ and every nonidentity term with $p/4$, where $p$ is the depolarization probability. For an error $E$ with row vector representation $r(E) = (x_1, \ldots, x_n, z_1, \ldots, z_n)$, we can count how many physical errors have occurred in total. Calling that number $i_1(E) \in \{0, \ldots, n\}$, the occurrence probability of $E$ is given by

$$\mathbb{P}_1(E) = \left(1 - \frac{3}{4}p\right)^{n-i_1(E)} \left(\frac{p}{4}\right)^{i_1(E)}. \tag{C1}$$

For two-qubit depolarization, we consider two individual errors $E_1 = (x_1, \ldots, x_n, z_1, \ldots, z_n)$ and $E_2 = (x_{n+1}, \ldots, x_{2n}, z_{n+1}, \ldots, z_{2n})$. Then the total error $E = (x_1, \ldots, x_{2n}, z_1, \ldots, z_{2n})$ has an occurrence probability of

$$\mathbb{P}_2(E) = \left(1 - \frac{15}{16}p\right)^{n-i_2(E)} \left(\frac{p}{16}\right)^{i_2(E)}, \tag{C2}$$

whereby $i_2(E) \in \{0, \ldots, n\}$. The integer $i_2(E)$ counts the number of times when at least one error has occurred on two physical qubits, so $n - i_2(E)$ is the number of double identities $I \otimes I$. As a simple example, consider two logical qubits, each composed of two physical qubits. Then $i_2(E) \in \{0, 1, 2\}$ and one needs to check whether $x_1, z_1, x_3, z_3$ are all zero or not (if not increase $i_2(E)$ by one) and whether $x_2, z_2, x_4, z_4$ are all zero or not.

For single-qubit dephasing, only the identity and the $Z$-operator are involved, so clearly, not all Pauli errors are possible anymore but only multiqubit phase flip errors which correspond to $x_i = 0$ for all $i \in \{0, \ldots, n\}$ and $z = (z_1, \ldots, z_n) \in \{0, 1\}^n$. The associated probability for such an error $E = (0, \ldots, 0, z_1, \ldots, z_n)$, where the number of physical errors is $i = \sum_{k=1}^{n} z_k$, is then given by

$$\mathbb{P}_3(E) = (1-p)^{n-i}p^i.$$

The run time here is a problem quickly encountered due to the exponentially increasing number of possible multiqubit Pauli errors when analyzing larger codes. What generally helps to reduce the run time significantly, is to work closer to the binary language that a computer uses anyway. First of all, one can translate the check representations of generators or operators into integers and, thereby, instead of comparing all entries of binary vectors, simply compare two numbers. Another example is to use the quick XOR operation on two integers instead of addition modulo two of two binary vectors or an OR operation, denoted by |, to quickly read out whether two values are zero. As an immediate application of the latter, we can express the number of physical single-qubit errors encountered previously in Eq. (C1) via

$$i_1(E) = \sum_{k=1}^{n} x_k | z_k$$

and the number of physical two-qubit errors (so operations different from $I \otimes I$) from Eq. (C2) via

$$i_2(E) = \sum_{k=1}^{n} x_k | z_k | x_{k+n} | z_{k+n}.$$

Furthermore, since symbolic computation is usually slow, it helps to delay any steps involving it as much as possible.

## APPENDIX D: EXPLICIT EXPRESSIONS FOR THE EFFECTIVE LOGICAL ERROR PARAMETERS

### 1. The three-qubit bit and phase flip codes

The three-qubit bit and phase flip codes have equal logical channels after depolarizing noise. For depolarizing single-qubit noise with depolarization probability $p$, the effective error parameters are given by

$$\lambda_X = \lambda_Y = \frac{5}{32}p^3 + \frac{3}{8}p^2\left(1 - \frac{3}{4}p\right)$$

and

$$\lambda_Z = \frac{1}{16}p^3 + \frac{3}{8}p^2\left(1 - \frac{3}{4}p\right) + \frac{3}{2}p\left(1 - \frac{3}{4}p\right)^2.$$

For two-qubit depolarization, it was found that the logical two-qubit channel has three nonoverlapping error curves, namely,

$$\lambda_{IX} = \lambda_{IY} = \lambda_{XI} = \lambda_{XZ} = \lambda_{YI} = \lambda_{YZ} = \lambda_{ZX} = \lambda_{ZY}$$
$$= \frac{55}{1024}p^3 + \frac{9}{64}p^2\left(1 - \frac{15}{16}p\right),$$

$$\lambda_{IZ} = \lambda_{ZI} = \lambda_{ZZ}$$
$$= \frac{5}{128}p^3 + \frac{21}{64}p^2\left(1 - \frac{15}{16}p\right) + \frac{3}{4}p\left(1 - \frac{15}{16}p\right)^2,$$

and

$$\lambda_{XX} = \lambda_{XY} = \lambda_{YX} = \lambda_{YY} = \frac{61}{1024}p^3 + \frac{3}{64}3p^2\left(1 - \frac{15}{16}p\right).$$

### 2. The five-qubit code

For single-qubit depolarization, the effective error parameters can be found to be

$$\lambda_X = \lambda_Y = \lambda_Z = \frac{33}{512}p^5 + \frac{45}{128}p^4\left(1 - \frac{3}{4}p\right)$$
$$+ \frac{35}{32}p^3\left(1 - \frac{3}{4}p\right)^2 + \frac{15}{8}p^2\left(1 - \frac{3}{4}p\right)^3,$$

whereby $p$ is the original depolarization probability. Thus, on the logical level, again depolarization occurs.

For two-qubit depolarization acting on the physical qubits constituting the logically encoded state, the new effective error parameters are given by

$$\lambda_{ZI} = \lambda_{YI} = \lambda_{XI} = \lambda_{IZ} = \lambda_{IY} = \lambda_{IX}$$
$$= \frac{23703}{524288}p^5 + \frac{7995}{32768}p^4\left(1 - \frac{15}{16}p\right)$$
$$+ \frac{965}{2048}p^3\left(1 - \frac{15}{16}p\right)^2 + \frac{105}{128}p^2\left(1 - \frac{15}{16}p\right)^3$$

and

$$\lambda_{ZX} = \lambda_{ZI} = \lambda_{ZZ} = \lambda_{YZ} = \lambda_{YY} = \lambda_{YX} = \lambda_{XZ}$$
$$= \lambda_{XX} = \lambda_{XY} = \frac{23763}{524288}p^5 + \frac{7815}{32768}p^4\left(1 - \frac{15}{16}p\right)$$
$$+ \frac{1145}{2048}p^3\left(1 - \frac{15}{16}p\right)^2 + \frac{45}{128}p^2\left(1 - \frac{15}{16}p\right)^3.$$

### 3. The Steane code

When the Steane code is subject to single-qubit dephasing, one gets a phase flip channel also on the logical level with

$$\lambda_Z = p^7 + 7p^6(1 - p) + 28p^4(1 - p)^3 + 7p^3(1 - p)^4$$
$$+ 21p^2(1 - p)^5.$$

When instead depolarizing noise affects single qubits, the effective error parameters read as

$$\lambda_X = \lambda_Y = \lambda_Z = \frac{575}{16384}p^7 + \frac{1225}{4096}p^6\left(1 - \frac{3}{4}p\right)$$
$$+ \frac{637}{512}p^5\left(1 - \frac{3}{4}p\right)^2 + \frac{371}{128}p^4\left(1 - \frac{3}{4}p\right)^3$$
$$+ \frac{231}{64}p^3\left(1 - \frac{3}{4}p\right)^4 + \frac{49}{16}p^2\left(1 - \frac{3}{4}p\right)^5.$$

With two-qubit depolarization, we obtain the logical error parameters

$$\lambda_{ZI} = \lambda_{YI} = \lambda_{XI} = \lambda_{IZ} = \lambda_{IY} = \lambda_{IX}$$
$$= \frac{10675163}{268435456}p^7 + \frac{4992997}{16777216}p^6\left(1 - \frac{15}{16}p\right)$$
$$+ \frac{494551}{524288}p^5\left(1 - \frac{15}{16}p\right)^2 + \frac{55685}{32768}p^4\left(1 - \frac{15}{16}p\right)^3$$
$$+ \frac{8199}{4096}p^3\left(1 - \frac{15}{16}p\right)^4 + \frac{385}{256}p^2\left(1 - \frac{15}{16}p\right)^5$$

and

$$\lambda_{ZX} = \lambda_{ZI} = \lambda_{ZZ} = \lambda_{YZ} = \lambda_{YY} = \lambda_{YX} = \lambda_{XZ} = \lambda_{XX} = \lambda_{XY}$$
$$= \frac{10685447}{268435456}p^7 + \frac{4964985}{16777216}p^6\left(1 - \frac{15}{16}p\right)$$
$$+ \frac{505843}{524288}p^5\left(1 - \frac{15}{16}p\right)^2 + \frac{54425}{32768}p^4\left(1 - \frac{15}{16}p\right)^3$$
$$+ \frac{6115}{4096}p^3\left(1 - \frac{15}{16}p\right)^4 + \frac{133}{256}p^2\left(1 - \frac{15}{16}p\right)^5.$$

### 4. The Shor code

Single-qubit dephasing acting on the nine-qubit Shor code results in a logical bit flip channel with an error probability given by

$$\lambda_X = p^9 + 9p^8(1 - p) + 9p^7(1 - p)^2 + 57p^6(1 - p)^3$$
$$+ 27p^5(1 - p)^4 + 99p^4(1 - p)^5 + 27p^3(1 - p)^6$$
$$+ 27p^2(1 - p)^7.$$

For single-qubit depolarization, the resulting logical channel has directed Pauli noise, namely

$$\lambda_X = \frac{4843}{262144}p^9 + \frac{14571}{65536}p^8\left(1 - \frac{3}{4}p\right)$$
$$+ \frac{20259}{16384}p^7\left(1 - \frac{3}{4}p\right)^2 + \frac{15387}{4096}p^6\left(1 - \frac{3}{4}p\right)^3$$

$$+ \frac{7065}{1024} p^5 \left(1 - \frac{3}{4}p\right)^4 + \frac{2601}{256} p^4 \left(1 - \frac{3}{4}p\right)^5$$

$$+ \frac{729}{64} p^3 \left(1 - \frac{3}{4}p\right)^6 + \frac{81}{16} p^2 \left(1 - \frac{3}{4}p\right)^7,$$

$$\lambda_Y = \frac{1447}{65536} p^9 + \frac{1791}{8192} p^8 \left(1 - \frac{3}{4}p\right)$$

$$+ \frac{4437}{4096} p^7 \left(1 - \frac{3}{4}p\right)^2 + \frac{987}{256} p^6 \left(1 - \frac{3}{4}p\right)^3$$

$$+ \frac{2313}{256} p^5 \left(1 - \frac{3}{4}p\right)^4 + \frac{315}{32} p^4 \left(1 - \frac{3}{4}p\right)^5$$

$$+ \frac{27}{16} p^3 \left(1 - \frac{3}{4}p\right)^6,$$

and

$$\lambda_Z = \frac{161}{8192} p^9 + \frac{3609}{16384} p^8 \left(1 - \frac{3}{4}p\right)$$

$$+ \frac{2439}{2048} p^7 \left(1 - \frac{3}{4}p\right)^2 + \frac{3903}{1024} p^6 \left(1 - \frac{3}{4}p\right)^3$$

$$+ \frac{477}{64} p^5 \left(1 - \frac{3}{4}p\right)^4 + \frac{639}{64} p^4 \left(1 - \frac{3}{4}p\right)^5$$

$$+ \frac{75}{8} p^3 \left(1 - \frac{3}{4}p\right)^6 + \frac{9}{4} p^2 \left(1 - \frac{3}{4}p\right)^7.$$

### 5. The nine-qubit surface code

When single-qubit dephasing acts on a logical surface code qubit, the resulting effective dephasing channel has a modified phase flip probability of

$$\lambda_Z = p^9 + 9p^8(1-p) + 18p^7(1-p)^2 + 28p^6(1-p)^3$$

$$+ 69p^5(1-p)^4 + 57p^4(1-p)^5 + 56p^3(1-p)^6$$

$$+ 18p^2(1-p)^7.$$

For single-qubit depolarization, the logical error parameters after error correction has been performed are

$$\lambda_X = \frac{2493}{131072} p^9 + \frac{7263}{32768} p^8 \left(1 - \frac{3}{4}p\right)$$

$$+ \frac{9945}{8192} p^7 \left(1 - \frac{3}{4}p\right)^2 + \frac{7731}{2048} p^6 \left(1 - \frac{3}{4}p\right)^3$$

$$+ \frac{3687}{512} p^5 \left(1 - \frac{3}{4}p\right)^4 + \frac{1293}{128} p^4 \left(1 - \frac{3}{4}p\right)^5$$

$$+ \frac{323}{32} p^3 \left(1 - \frac{3}{4}p\right)^6 + \frac{33}{8} p^2 \left(1 - \frac{3}{4}p\right)^7,$$

$$\lambda_Y = \frac{1397}{65536} p^9 + \frac{1791}{8192} p^8 \left(1 - \frac{3}{4}p\right)$$

$$+ \frac{4587}{4096} p^7 \left(1 - \frac{3}{4}p\right)^2 + \frac{987}{256} p^6 \left(1 - \frac{3}{4}p\right)^3$$

$$+ \frac{2163}{256} p^5 \left(1 - \frac{3}{4}p\right)^4 + \frac{315}{32} p^4 \left(1 - \frac{3}{4}p\right)^5$$

$$+ \frac{77}{16} p^3 \left(1 - \frac{3}{4}p\right)^6,$$

and

$$\lambda_Z = \frac{1249}{65536} p^9 + \frac{1815}{8192} p^8 \left(1 - \frac{3}{4}p\right)$$

$$+ \frac{4967}{4096} p^7 \left(1 - \frac{3}{4}p\right)^2 + \frac{967}{256} p^6 \left(1 - \frac{3}{4}p\right)^3$$

$$+ \frac{1847}{256} p^5 \left(1 - \frac{3}{4}p\right)^4 + \frac{323}{32} p^4 \left(1 - \frac{3}{4}p\right)^5$$

$$+ \frac{161}{16} p^3 \left(1 - \frac{3}{4}p\right)^6 + 4p^2 \left(1 - \frac{3}{4}p\right)^7.$$

### 6. The eleven-qubit code

Using adaptive error syndrome identification, when the eleven-qubit code is subject to single-qubit dephasing, we obtain a phase flip channel with the modified error probability

$$\lambda_Z = p^{11} + 11p^{10}(1-p) + 55p^9(1-p)^2 + 146p^8(1-p)^3$$

$$+ 202p^7(1-p)^4 + 263p^6(1-p)^5 + 199p^5(1-p)^6$$

$$+ 128p^4(1-p)^7 + 19p^3(1-p)^8.$$

In the case of single-qubit depolarization affecting the physical qubits of the eleven-qubit code, the logical error probabilities are given by

$$\lambda_X = \frac{22031}{2097152} p^{11} + \frac{159895}{1048576} p^{10} \left(1 - \frac{3}{4}p\right)$$

$$+ \frac{535}{512} p^9 \left(1 - \frac{3}{4}p\right)^2 + \frac{17169}{4096} p^8 \left(1 - \frac{3}{4}p\right)^3$$

$$+ \frac{44055}{4096} p^7 \left(1 - \frac{3}{4}p\right)^4 + \frac{40923}{2048} p^6 \left(1 - \frac{3}{4}p\right)^5$$

$$+ \frac{1829}{64} p^5 \left(1 - \frac{3}{4}p\right)^6 + \frac{913}{32} p^4 \left(1 - \frac{3}{4}p\right)^7$$

$$+ \frac{667}{32} p^3 \left(1 - \frac{3}{4}p\right)^8 + \frac{27}{16} p^2 \left(1 - \frac{3}{4}p\right)^9,$$

$$\lambda_Y = \frac{11545}{1048576} p^{11} + \frac{80147}{524288} p^{10} \left(1 - \frac{3}{4}p\right)$$

$$+ \frac{134191}{131072} p^9 \left(1 - \frac{3}{4}p\right)^2 + \frac{136551}{32768} p^8 \left(1 - \frac{3}{4}p\right)^3$$

$$+ \frac{90279}{8192} p^7 \left(1 - \frac{3}{4}p\right)^4 + \frac{41913}{2048} p^6 \left(1 - \frac{3}{4}p\right)^5$$

$$+ \frac{14365}{512} p^5 \left(1 - \frac{3}{4}p\right)^6 + \frac{3277}{128} p^4 \left(1 - \frac{3}{4}p\right)^7$$

$$+ \frac{475}{32} p^3 \left(1 - \frac{3}{4}p\right)^8,$$

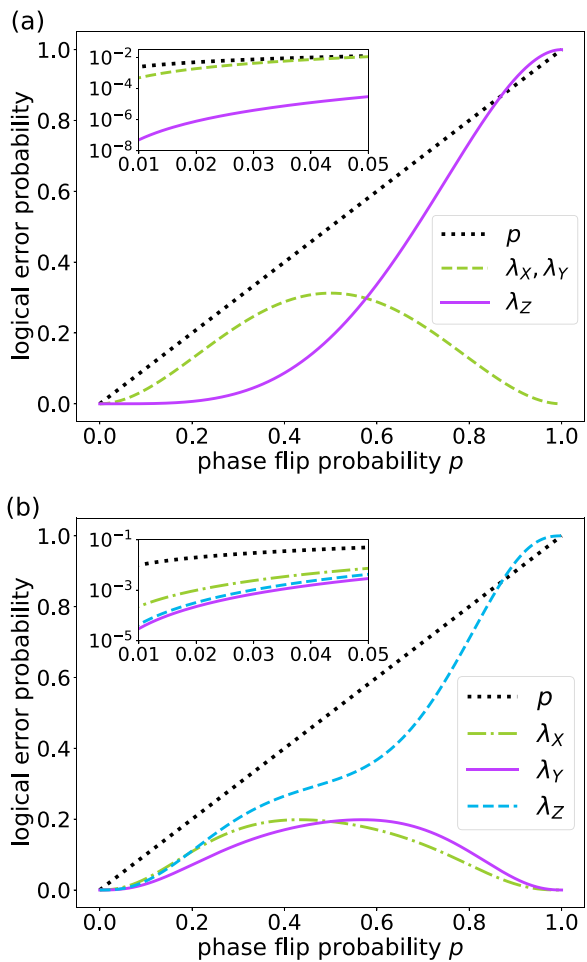

FIG. 17. Logical error parameters with standard, nonadaptive error syndrome identification against single-qubit dephasing (a) for the five-qubit code and (b) the eleven-qubit code.

$$\lambda_Z = \frac{11547}{1048576}p^{11} + \frac{80139}{524288}p^{10}\left(1-\frac{3}{4}p\right)$$
$$+ \frac{134183}{131072}p^9\left(1-\frac{3}{4}p\right)^2 + \frac{136575}{32768}p^8\left(1-\frac{3}{4}p\right)^3$$
$$+ \frac{90279}{8192}p^7\left(1-\frac{3}{4}p\right)^4 + \frac{41889}{2048}p^6\left(1-\frac{3}{4}p\right)^5$$

$$+ \frac{14373}{512}p^5\left(1-\frac{3}{4}p\right)^6 + \frac{3285}{128}p^4\left(1-\frac{3}{4}p\right)^7$$
$$+ \frac{471}{32}p^3\left(1-\frac{3}{4}p\right)^8.$$

## APPENDIX E: NONADAPTIVE ERROR SYNDROME IDENTIFICATION

For the five-qubit and eleven-qubit codes, the adaptive error syndrome identification allows to obtain logical dephasing channels with improved phase flip probabilities for low noise. If one instead uses the standard error syndrome identification, mapping error syndromes to minimal weight-Pauli errors, the resulting logical channel after single-qubit dephasing has nonzero $\lambda_X$ and $\lambda_Y$, see Fig. 17.

The error parameters of the five-qubit code in Fig. 17(a) are then given by

$$\lambda_X = \lambda_Y = 5p^3(1-p)^2 + 5p^2(1-p)^3,$$
$$\lambda_Z = p^5 + 5p^4(1-p).$$

For single-qubit dephasing acting on the eleven-qubit code, Fig. 17(b), the resulting logical channel has the error parameters

$$\lambda_X = 32p^8(1-p)^3 + 53p^7(1-p)^4 + 96p^6(1-p)^5$$
$$+ 92p^5(1-p)^6 + 74p^4(1-p)^7 + 47p^3(1-p)^8$$
$$+ 2p^2(1-p)^9,$$

$$\lambda_Y = 2p^9(1-p)^2 + 47p^8(1-p)^3 + 74p^7(1-p)^4$$
$$+ 92p^6(1-p)^5 + 96p^5(1-p)^6 + 53p^4(1-p)^7$$
$$+ 32p^3(1-p)^8,$$

$$\lambda_Z = p^{11} + 11p^{10}(1-p) + 53p^9(1-p)^2 + 40p^8(1-p)^3$$
$$+ 107p^7(1-p)^4 + 113p^6(1-p)^5 + 161p^5(1-p)^6$$
$$+ 96p^4(1-p)^7 + 46p^3(1-p)^8.$$